\documentclass[conference]{IEEEtran}
\ifCLASSINFOpdf
\else
\fi
\usepackage{url}
\usepackage{soul}
\usepackage{enumitem}
\usepackage{color}
\usepackage{hyperref}
\usepackage{
  color,
  float,
  epsfig,
  wrapfig,
  graphics,
  graphicx,
  subcaption
}
\usepackage{tcolorbox}
\usepackage{pifont}
\usepackage{booktabs}

\definecolor{pinkcolor}{HTML}{F19C99}
\definecolor{orangecolor}{HTML}{FFB570}
\definecolor{yellowcolor}{HTML}{FFD966}
\definecolor{tealcolor}{HTML}{67AB9F}
\definecolor{purplecolor}{HTML}{C3ABD0}
\definecolor{greencolor}{HTML}{80A468}
\definecolor{blackcolor}{HTML}{000000}

\newcommand{\blackcapsule}[1]{\tikz[baseline=(c.base)]{
    \node[draw, rounded corners, inner sep=2pt, fill=black, text=white] (c) {\footnotesize#1};
}}

\usepackage{longtable}
\usepackage{booktabs}
\usepackage{array}
\usepackage{ragged2e}
\usepackage{xurl}      
\usepackage{hyperref}  

\newcolumntype{L}[1]{>{\RaggedRight\arraybackslash}p{#1}}

\newcommand{\cbtheme}[2]{%
  \multirow{#1}{*}{%
    \rotatebox[origin=c]{90}{%
      \parbox{3.2cm}{\centering\footnotesize #2}%
    }%
  }%
}
\usepackage{graphicx}
\usepackage{tabularx}
\usepackage{multirow}
\usepackage{multicol}
\usepackage{booktabs}
\usepackage{pifont}
\usepackage{xcolor}
\usepackage{comment}
\newcommand{\cmark}{\textcolor{green!60!black}{\ding{51}}}
\newcommand{\xmark}{\textcolor{red}{\ding{55}}}
\newcommand{\lmark}{\textcolor{orange!90!black}{\ding{55}}{\footnotesize\,(Limited)}}

\begin{document}
%
\title{From Frontier to Shadow AI: A Simmering Threat to Assurance and Security in Critical Infrastructure}

\author{
\IEEEauthorblockN{
Mohan Baruwal Chhetri, 
Shahroz Tariq, 
Tooba Aamir, 
Marthie Grobler, 
Chandra Thapa, 
Ronal Singh
}
\IEEEauthorblockA{
CSIRO, Australia \\
}
\IEEEauthorblockA{
\{mohan.baruwalchhetri, shahroz.tariq, tooba.aamir, marthie.grobler, chandra.thapa, ronal.singh\}@csiro.au
}
}

\maketitle

\begin{abstract}

Frontier AI systems, including large language models and emerging agentic AI tools, offer significant operational benefits but present unique challenges to critical infrastructure (CI) environments due to their non-deterministic and emergent properties. While formal adoption is inherently cautious and tightly controlled due to strict regulatory oversight, widespread accessibility has catalysed \textit{shadow AI}: the unsanctioned use of frontier AI outside established organisational controls. In CI settings, shadow AI bypasses established assurance and oversight mechanisms, amplifying risks to data protection, decision reliability, and regulatory compliance, with potential consequences for essential service delivery. We present the first empirical study of shadow AI in CI environments, characterising it as a systemic socio-technical condition of assurance erosion. Drawing on semi-structured interviews with senior executives and functional leaders across 27 Australian CI organisations (\textit{Communications}, \textit{Energy}, and \textit{Water and Sewerage} sectors), we analyse how shadow AI manifests in practice, how it interacts with existing technical and governance controls, and the resulting security, assurance, and compliance risks. We develop an empirically derived threat model identifying three primary mechanisms of security degradation: (i) \textit{boundary bypass}, where data flows circumvent established perimeters; (ii) \textit{unassessed capability expansion}, where embedded AI features introduce latent risks; and (iii) \textit{loss of observability} via \textit{governance circumvention}, undermining forensic auditability and least-privilege enforcement. Our findings demonstrate that shadow AI introduces unmanaged risks that fundamentally challenge existing security and compliance frameworks, necessitating tailored, pathway-aligned governance and control strategies. 

\end{abstract}




%
\IEEEpeerreviewmaketitle

\section{Introduction} \label{sec:intro}

\textbf{Frontier AI} refers to advanced, general-purpose AI systems operating at the limits of current capability in reasoning, coding, and multimodal understanding. This includes foundation models and agentic AI systems capable of autonomous, goal-directed planning and coordinated execution across heterogeneous tools and systems~\cite{uk_dsit_frontier_ai_2025}. These systems offer substantial efficiency gains and improved decision-making~\cite{Luo2026ShadowCache,bommasani2021opportunities}, particularly in environments characterised by scale, complexity, and time-critical demands. However, their generality, autonomy, and non-deterministic and emergent behaviours strain existing security, assurance, and governance regimes, introducing risks including diminished human oversight, opaque or unreliable decision-making, unmonitored data exposure, and emergent autonomy in high-stakes contexts~\cite{uk_dsit_frontier_ai_2025,mitch2025governance}. Collectively, these properties distinguish frontier AI from earlier task-specific or rule-based, largely deterministic systems and reshape expectations for governance, oversight, and assurance. 

The low-cost, low-friction accessibility of frontier AI systems is accelerating the emergence of \textbf{shadow AI}: informal and unsanctioned use outside organisational controls, often driven by productivity pressures, capability gaps, or unclear organisational guidance~\cite{kshetri2025transforming, cerchione2026artificial, van2024bring, nrg2024outoftheshadows}. Unlike sanctioned AI deployments, shadow AI operates with minimal institutional visibility, bypassing security, assurance, and compliance checks. In security terms, shadow AI constitutes a socio‑technical threat condition in which authorised insiders, typically without malicious intent, can introduce opaque data flows, undocumented third‑party dependencies, and unlogged AI‑mediated actions into core organisational workflows~\cite{matsumoto2025a2a, puthal2025shadows, sikos2025securing, kwan2024navigating}. These practices can undermine core security and assurance properties, 
reframing the threat model from external adversaries and malicious insiders to non-malicious, insider-driven, AI-mediated exposure pathways.

\textbf{Critical infrastructure (CI)}\footnote{CI comprises the systems, assets, facilities, and networks that deliver essential services and underpin national security, economic security, prosperity, and public health and safety~\cite{publicsafetycanada2014,cisc2023_space}.} is a tightly regulated, safety-critical domain where shadow AI introduces significant security-relevant challenges. This arises from the defining characteristics of CI, including severe consequences of failure, systemic interdependence, and limited tolerance for disruption, which collectively motivate strong regulatory oversight~\cite{nsm2024}. Across jurisdictions, CI operators are subject to prescriptive regulatory frameworks that impose positive security obligations, reinforce governance and accountability requirements, and mandate incident reporting. For example, Australia's \textit{Security of Critical Infrastructure (SOCI) Act}~\cite{soci2018}, alongside comparable regimes in the United Kingdom (UK), European Union (EU), and United States~(US) \cite{nsm2024, csrbill2025, eu2022nis2, circia_2022}, reflects a broader international shift toward enforceable obligations for security, resilience, and proactive risk management.
Under these frameworks, CI governance typically presumes clearly defined system boundaries, observable system behaviour, and identifiable accountability structures to enable compliance, assurance, and evidentiary reporting~\cite{cirmp2023}. 

Given this regulatory context, formal frontier AI adoption in CI environments remains cautious and tightly controlled, aligned with national security guidance and evolving regulatory expectations~\cite{asd2025aiot, Lee2025MOdeling, slay2026socireview}. In contrast, shadow AI reflects a less regulated and structurally misaligned practice that departs from the visibility- and assurance-centric practices essential to CI operations. By bypassing approved monitoring and assurance processes, shadow AI creates conditions that introduce gaps in visibility, attribution, and provenance, which can complicate auditability and post‑incident reconstruction~\cite{Silic2025FromShadow}, both of which are foundational to security governance in regulated CI environments. 

These conditions reduce control effectiveness by bypassing data-flow controls (\textit{confidentiality}), introducing unvalidated and non-deterministic transformations into operational artefacts (\textit{integrity}), and eroding the provenance records required for \textit{auditability} and post‑incident recovery. These effects do not necessarily result in immediate failures but limit the ability to monitor, verify, and enforce security properties in practice. These dynamics are amplified by the tightly coupled and interdependent nature of CI systems, where localised failures can cascade across organisational, sectoral, or national boundaries~\cite{Lee2025MOdeling,slay2026socireview}. Together, they create a misalignment between formal governance assumptions (clearly bounded systems, observable behaviour, and accountable operations), and AI-mediated practices that are informal, distributed, and only partially visible. We conceptualise this phenomenon as \textit{assurance erosion}. This misalignment weakens the conditions required for auditability, assurance, and evidentiary reporting, creating compliance gaps in which organisations may appear compliant while lacking the artefacts needed to demonstrate effective oversight in practice~\cite{Edwards2023AISociety}.

Despite these implications, empirical evidence on the prevalence, forms, and handling of shadow AI in CI settings remains limited~\cite{bommasani2021opportunities, Mansner2025ShadowAI}. Prior work has largely examined it as a general organisational or productivity phenomenon, with little direct focus on CI environments~\cite{Silic2025FromShadow}, constraining understanding of how shadow AI use emerges in CI operations, and how its security, governance, and compliance risks are managed in practice. 
%
To address this gap, we present, to our knowledge, the first CI-focused empirical analysis of shadow AI, situated within a broader study of frontier AI adoption in CI. We examine how its use is shaped by, and interacts with, security, governance, and assurance dynamics, alongside CI-specific regulatory obligations.
We investigate shadow AI in CI organisations through the following research questions:
\begin{itemize}
    \item \textbf{RQ1:} How does the informal or unsanctioned use of frontier AI manifest within CI organisations, and what patterns of use, tools, and workflows shape its incorporation into everyday work practices?
    \item \textbf{RQ2:} How do existing technical, organisational, and governance controls interact with observed shadow AI practices, and where do breakdowns or blind spots allow such use to persist beyond formal assurance processes?
    \item \textbf{RQ3:} What security-relevant risks emerge from observed patterns of shadow AI use, and how do these risks persist, compound, or remain latent over time?
\end{itemize}

Drawing on semi‑structured interviews with senior executives and functional leaders across 27 Australian CI operators, this study empirically examines shadow AI as a security, governance, and assurance challenge in regulated environments\footnote{We focus on Australia as a representative jurisdiction given its mature CI regulatory framework, with established compliance and assurance cycles and active supervisory and enforcement mechanisms~\cite{slay2026socireview}.}. We focus on the \textit{Communications}, \textit{Energy}, and \textit{Water and Sewerage} sectors, selected for their essential service delivery, role as key cross-sector dependencies, and documented exposure to escalating threat activity~\cite{DiasBaptista2026Interdependencies,slay2026socireview}. From this empirical base, we derive a threat model that frames shadow AI not as isolated policy non-compliance, but as a systemic socio-technical condition shaped by human practices, organisational constraints, and regulatory context. By establishing shadow AI as a first-order risk in CI environments, this work extends security research on CI protection and demonstrates that traditional governance, assurance, and control mechanisms are limited in their ability to address informal use of frontier AI. While no major incidents were observed, our findings reveal a pervasive undercurrent of shadow AI use that signals elevated risk of future security, governance, or compliance failures.

\textbf{Our contributions include:} (i) an empirical characterisation of shadow AI practices in CI; (ii) a structured analysis of how shadow AI interacts with existing technical, organisational, and governance controls in CI; and 
(iii) a socio-technical analysis of security, governance, and assurance risks, introducing the concept of \textit{assurance erosion} as a unifying framework that explains how shadow AI systematically degrades CI security, auditability, and compliance guarantees. Together, these contributions characterise shadow AI as a systemic socio-technical risk in CI. 
\section{Shadow AI in CI: Regulatory, Governance, and Assurance Misalignment} 
\label{sec:background}

\subsection{Critical Infrastructure: Scope and Systemic Characteristics} 

CI refers to physical and digital systems that provide essential services, the disruption or degradation of which would have severe consequences for national security, economic stability, or public health and safety~\cite{soci2018, cisa2024sectors}. Examples include energy generation and distribution, communications, and water and wastewater services. CI systems are highly interdependent and increasingly digitalised, amplifying systemic risk such that attacks on, or disruptions to, a single asset in one sector can cascade rapidly across sectors \cite{slay2026socireview,DiasBaptista2026Interdependencies,oecd2019}.


Historically, CI protection emphasised physical security and redundancy~\cite{ghosh2010cip_history,orourke2007ci_resilience}. However, IT--OT convergence, supply-chain dependence, and escalating cyber-enabled disruptions have exposed the limits of voluntary security practices for managing systemic risk~\cite{gao2023cisa}. In response, governments have adopted outcome-oriented regulatory frameworks that impose positive security obligations, mandatory incident reporting, and enhanced governance and accountability requirements for CI owners and operators~\cite{csrbill2025,eu2022nis2,nsm2024,soci2018,circia_2022}. These frameworks embed implicit assumptions about system visibility, controllability, and enforceable compliance, which underpin contemporary CI security and assurance practice.

\subsection{Regulatory Frameworks and Assurance Obligations}

Across major jurisdictions, CI regulation has converged on risk-based governance models that prioritise visibility, accountability, and post-incident assurance. In Australia, the \textit{SOCI Act~2018}~\cite{soci2018} establishes a technology-neutral framework spanning 11 CI sectors, requiring asset registration, mandatory cyber incident reporting, and outcome-oriented risk management obligations. Responsible entities must maintain a Critical Infrastructure Risk Management Program (CIRMP) covering cyber and information security, supply-chain dependencies, personnel, and physical hazards~\cite{cirmp2023}. 

Similar regulatory shifts are evident globally. In the US, the National Security Memorandum on Critical Infrastructure Security and Resilience (\textit{NSM-22})~\cite{nsm2024} empowers \textit{Sector Risk Management Agencies (SRMAs)} to define and enforce sector-specific cyber security requirements. In the UK, reforms to the Security of Network \& Information Systems Regulations (\textit{NIS Regulations})~\cite{nis2018}, including the proposed \textit{Cyber Security and Resilience Bill}, strengthen oversight and senior-level accountability~\cite{csrbill2025}, while the EU's \textit{NIS2 Directive} expands regulatory scope and incident-reporting obligations across critical sectors~\cite{eu2022nis2}. Collectively, these regimes reflect a shift toward enforceable assurance, supply-chain risk management, and outcome-based accountability. 

\subsection{Breakdown of Regulatory and Assurance Assumptions}

Shadow AI challenges core assumptions embedded in contemporary CI regulatory and assurance frameworks that underpin confidence in CI security, assurance, and compliance~\cite{cirmp2023}. These frameworks assume that operational systems and their interdependencies are identifiable, bounded\footnote{In this context, ``bounded'' refers to systems and dependencies with clearly defined scope, ownership, and governance, such that their operation, risks, and controls can be identified, monitored, and assured.}, and visible to governance and assurance processes; that risk-bearing activities are observable through approved systems and monitoring mechanisms; and that controls, mitigations, and system behaviour can be demonstrated and reconstructed through audit and post-incident analysis. They further presume that responsibility for system behaviour and consequential decisions can be clearly attributed to defined roles and governance structures, and that compliance and risk-management artefacts accurately reflect actual operational practices.



\subsection{Shadow AI as a Governance and Assurance Problem}

Shadow AI is inherently misaligned with prevailing governance and assurance models. CI regulatory frameworks assume that risk-bearing capabilities are deliberately introduced, centrally governed, and integrated within bounded systems that are subject to defined oversight and assurance. By contrast, shadow AI enables frontier AI capabilities to be informally and incrementally adopted within everyday workflows, bypassing the points where security controls, assurance, auditability, and accountability are expected to apply. This structural misalignment reduces operational visibility and auditability, weakens evidence-based assurance and accountability, and undermines the foundations of security governance in safety-critical CI environments~\cite{Silic2025FromShadow}. It therefore motivates empirical investigation into how shadow AI manifests in CI operations and how its risks are recognised and managed in practice.

\section{Research Design and Methodology} \label{sec:methodology}

We employ a qualitative, interview‑based approach to examine shadow AI as a security and governance phenomenon that cannot be meaningfully observed through logs, surveys, or incident data alone.
Using a socio‑technical lens, we analyse how informal frontier AI emerges from interactions between organisational practices, decision-making structures, and outcome-oriented regulatory and assurance regimes. 

\subsection{Interview Design}

We conducted semi‑structured interviews with executives, senior managers, and functional leaders responsible for organisational responses to AI risk and compliance in CI settings. Interview questions were designed to elicit organisational-level perspectives on informal or unsanctioned AI use, focusing on how such practices are perceived and governed within CI-owning or operating organisations.

Interview prompts focused on four areas: (i) organisational awareness and perceptions of shadow AI; (ii) observed or suspected informal AI use and its operational contexts; (iii) perceived security, safety, and regulatory risks arising from such use; and (iv) existing or proposed organisational measures for governing AI, including detection, mitigation, and assurance. 
This design supported discussion of shadow AI across strategic, governance, and operational levels, while minimising framing that could inhibit candid organisational reporting. 

\subsection{Recruitment}

Participants were recruited from three CI sectors: \textit{Communications}, \textit{Energy}, and \textit{Water and Sewerage}, which underpin Australia's CI ecosystem and provide relevant contexts for examining the governance, security, and assurance implications of shadow AI \cite{DiasBaptista2026Interdependencies}. 
We focused on senior leaders because shadow AI constitutes a governance and assurance failure mode rather than a purely individual behaviour; these roles are uniquely positioned to observe how informal AI use intersects with enterprise governance structures, security controls, regulatory obligations, and incident‑response processes.
Participants were recruited through established sector-specific and professional networks across government and industry. This enabled access to experienced practitioners while preserving trust, confidentiality, and sector sensitivity. No financial or other incentives were offered for participation.

\subsection{Data Collection and Analysis}

Interviews were conducted between December 2025 and April 2026, each lasting 45--60 minutes. All 27 interviews were held via secure Microsoft Teams and facilitated by two members of the research team. With participant consent, interviews were recorded, professionally transcribed, and subsequently de-identified to remove organisational and personal identifiers before analysis. Interview data were analysed using a hybrid qualitative approach that combined structured extraction and reflexive thematic analysis. Structured extraction captured factual and categorical information to support transparent cross-case comparison, while interpretive material was thematically analysed to examine how participants understood shadow AI, associated risks, and governance practices.

Consistent with reflexive thematic analysis, coding was treated as an interpretive and iterative process rather than a mechanical exercise aimed at maximising coder agreement~\cite{braun2021}. Three researchers jointly coded an initial subset of six interviews to develop and calibrate a shared analytic framework. The remaining interviews were coded independently, with regular team discussions to compare interpretations, refine code definitions, and iteratively develop the thematic structure. Analytic coherence was achieved through reflexive dialogue and documented consensus rather than inter-coder reliability metrics, with an audit trail of codebook revisions and analytic memos maintained to support trustworthiness~\cite{nowell2017thematic}.

\subsection{Limitations}

As with all interview-based qualitative research, this study has two inherent limitations. First, it focuses on three CI sectors: \textit{Communications}, \textit{Energy}, and \textit{Water and Sewerage}, within the SOCI regulatory context. While these sectors represent tightly coupled, high-consequence infrastructure, shadow AI practices and governance dynamics may differ in other CI sectors, such as \textit{Healthcare and Medical} or \textit{Financial Services and Markets}. At the same time, the Australian setting provides a mature, outcome-oriented regulatory baseline that enables systematic analysis and offers a useful point of comparison for other CI regimes. 

Second, the findings are based on self-reported accounts and therefore reflect partial organisational visibility and institutional framing of risks and controls. Focusing on senior leadership, strategic, governance, and operational roles provides a strong enterprise-level view on how shadow AI intersects with formal controls and regulatory obligations. While this may under-represent highly localised frontline practices, cross-role and cross-sector sampling helps mitigate this limitation. 

This study is designed for analytic generalisation rather than statistical representativeness. By sampling roles responsible for security, digital, governance, and AI decisions, we prioritise visibility into organisation-level controls, accountability structures, and assurance processes, while acknowledging that some highly localised user practices may remain under-observed.

Despite these limitations, the study provides empirically grounded insights into an under-examined phenomenon in regulated, safety-critical environments and 
offers a foundation for future comparative, longitudinal, or mixed-methods research on shadow AI across sectors and jurisdictions.

\section{Results} \label{sec:results}


\subsection{Demographics}

We conducted interviews with 32 participants representing 27 organisations, spanning executive, senior management, and leadership roles in digital, data and AI, governance, security, architecture, and AI enablement. Organisations represented both public and private sectors and varied in size, from small enterprises ($\leq$ 100 employees) to large organisations ($\geq$~5,000 employees). To preserve anonymity, \autoref{tab:demographics} reports aggregated demographics. Most interviews involved a single participant, with three interviews involving two participants and one involving three. All interviews were conducted and analysed at the organisational level. 

\begin{table*}[t]
\centering
\caption{Participant Demographics}
\label{tab:demographics}
\footnotesize
\setlength{\tabcolsep}{4pt}
\renewcommand{\arraystretch}{0.95}
\begin{tabular*}{\textwidth}{@{\extracolsep{\fill}}lrlrlrlrlrlr@{}}
\toprule
\multicolumn{2}{c}{\textbf{Sector}} &
\multicolumn{2}{c}{\textbf{Org. Type}} &
\multicolumn{2}{c}{\textbf{Employees}} &
\multicolumn{2}{c}{\textbf{Gender}} &
\multicolumn{2}{c}{\textbf{Education}} &
\multicolumn{2}{c}{\textbf{Job Role}} \\
\cmidrule(lr){1-2}
\cmidrule(lr){3-4}
\cmidrule(lr){5-6}
\cmidrule(lr){7-8}
\cmidrule(lr){9-10}
\cmidrule(l){11-12}
\textit{Category} & \textit{n} &
\textit{Category} & \textit{n} &
\textit{Category} & \textit{n} &
\textit{Category} & \textit{n} &
\textit{Category} & \textit{n} &
\textit{Category} & \textit{n} \\
\midrule
Water \& Sewerage & 10 & Public Sector     & 13 & $\leq$100  & 3  & Female & 10 & PhD       & 1  & Executive Leadership           & 6  \\
Energy            & 10 & Private Sector    & 10 & 101--1000  & 10 & Male   & 22 & Masters   & 10 & Senior Management              & 12 \\
Communications    & 7  & Public--Private   & 2  & 1001--5000 & 12 &        &    & Bachelors & 17 & Digital, Data \& AI Leadership & 7  \\
                  &    & Not-For-Profit    & 2  & $\geq$5000 & 2  &        &    & Other     & 4  & Governance \& Policy           & 2  \\
                  &    &                   &    &            &    &        &    &           &    & Architecture / IT Ops          & 2  \\
                  &    &                   &    &            &    &        &    &           &    & AI Enablement                  & 2  \\
                  &    &                   &    &            &    &        &    &           &    & Security \& Risk               & 1  \\
\bottomrule
\end{tabular*}
\vspace{-10pt}
\end{table*}

\subsection{Prevalence of Shadow AI}

Across the 27 organisations covered, participants consistently described shadow AI as widespread, characterised by routine informal use of public generative AI tools. This use was reported as predominantly interactive and human-in-the-loop, with no evidence of agentic or safety-critical deployment. 

\subsubsection{Overall Prevalence} Only one organisation reported no observed shadow AI usage, attributing this to early, comprehensive blocking of non-approved AI tools (\blackcapsule{P04}). In contrast, participants across all other organisations described shadow AI as common and widespread rather than exceptional, with some estimating usage at \textit{``close to 80 or 90\% of people''} (\blackcapsule{P14}) 
Others expressed strong confidence in its presence, stating \textit{``I'm 100\% sure that there is use of shadow AI in the organisation''} (\blackcapsule{P20}), \textit{``shadow AI is extremely prevalent''} (\blackcapsule{P15}), and \textit{``anyone that says no to that is probably dreaming''} (\blackcapsule{P18}, while others noted organisational norms that allow experimentation under broad constraints (\blackcapsule{P26}). 

\subsubsection{Organisational Spread} Participants described shadow AI use as organisation-wide rather than confined to specific teams, functions, or roles, with use reported among operational staff, office-based employees, developers, managers, and privileged technical roles. One participant noted, \textit{``It's across the organisation''} (\blackcapsule{P18}), while another acknowledged personally using shadow AI to prepare a technical report (\blackcapsule{P25}). Others observed that administrators with elevated permissions were also engaging in informal AI use (e.g., \blackcapsule{P11}, \blackcapsule{P25}). Hybrid and remote working arrangements further diffuse and obscure usage, with one participant noting \textit{``There's plenty of people working from home \dots  people out on site \dots''} (\blackcapsule{P13}) potentially using shadow AI on personal devices. 

\subsubsection{Persistence Beyond Sanctioned Tools} Participants consistently described shadow AI as predating sanctioned tools and persisting after official alternatives were deployed. In several organisations, informal practices were already entrenched by the time internal policies or approved tools were introduced. One participant explained, \textit{``By the time we put those policies in place, people were already using ChatGPT \dots the inertia of that has continued''} (\blackcapsule{P05}). Although sanctioned tools were seen as reducing risk, they co-exist with shadow use rather than fully displacing it, particularly where users perceived gaps in functionality or accessibility (e.g., \blackcapsule{P18},~\blackcapsule{P24}).

\subsubsection{Absence of Shadow Agentic AI} While informal AI use was widespread, participants reported no evidence of shadow \emph{agentic} AI, describing usage as primarily interactive, chat-based, and human-in-the-loop, rather than involving advanced automation or system integration. Several explicitly rejected the presence of covert autonomous AI activity, describing usage as \textit{``very casual shadow AI''}, emphasising that \textit{``It's just chat, no one is building giant LLMs, agentic agents, \dots in the dark''} (\blackcapsule{P08}); others similarly stated \textit{``I don't believe we have any shadow agentic AI''} (\blackcapsule{P17}) and that \textit{``Crew AI and agent platforms \dots we're not seeing any of that at all''} (\blackcapsule{P02}). 

\begin{tcolorbox}[
  colback=gray!5,
  colframe=black!40,
  boxrule=0.6pt,
  rounded corners,
  boxsep=2pt,
  left=1.5mm,
  right=1.5mm,
  top=0.5mm,
  bottom=0.5mm
]
{\textbf{\ul{Key Insight:}} 
\textit{Shadow AI is highly prevalent and routinely embedded in everyday work; however, observed use is largely limited to interactive, human-in-the-loop LLM interactions, with no evidence of shadow agentic AI.}}


\end{tcolorbox}

\subsection{Drivers of Shadow AI} 

Participants identified a consistent set of organisational, cultural, and structural factors driving the emergence and persistence of shadow AI, framing it as a practical, and often rational, response to everyday work demands rather than isolated non-compliance.

\subsubsection{Productivity Pressures and Recognised Efficiency Gains} Participants consistently described productivity gains as a primary driver of shadow AI, particularly for text- and knowledge-intensive tasks. AI tools were seen to save time on drafting, summarisation, and ideation, and were often framed as pragmatic support for routine work. Participants commented on its ability to streamline time-consuming tasks (\blackcapsule{P06}) and support meeting delivery timelines (\blackcapsule{P27}). Delays in deploying sanctioned tools were described as sustaining informal use, with some organisations tolerating shadow AI as a temporary productivity measure, reflected in statements such as \textit{``we'll probably block it after we've given them a corporate-sanctioned version''} (\blackcapsule{P08}). 

\subsubsection{Ease of Access and Low Barriers to Adoption} Participants consistently highlighted the low barriers to accessing public AI tools, often available at little or no personal cost and readily accessible across work and personal devices. Tools such as ChatGPT, Gemini, and Claude were described as easy to adopt without organisational approval, as reflected in comments such as \textit{``It's so easy right now to procure a licence''} (\blackcapsule{P12}) and \textit{``Everybody has Gemini on their phone''} (\blackcapsule{P27}). The cross-device availability, including \textit{``on their phone \dots their iPad \dots and their computer''} (\blackcapsule{P08}), made informal adoption difficult to resist, particularly when sanctioned alternatives were perceived as slower to deploy or functionally constrained. 

\subsubsection{Executive Signalling and Permission to Experiment} Several participants described organisational cultures where executive enthusiasm for AI innovation implicitly permitted experimentation, even where formal governance frameworks lagged. Leadership interest was interpreted as encouraging exploration, as reflected in comments such as \textit{``Some people are using it anyway \dots''} (\blackcapsule{P01}) and leadership being \textit{``open to the application \dots experimentation \dots user experience''} (\blackcapsule{P23}), lowering social and psychological barriers to informal AI use. Participants stressed that such use was rarely malicious, instead reflecting efforts to work effectively amid ambiguous or evolving rules. 

\subsubsection{Infrastructural Limitations and Fragmented Data} Weaknesses in internal digital infrastructure and constrained availability of sanctioned tools were also identified as drivers of Shadow AI use. Legacy databases, fragmented systems, and poorly integrated data environments were described as hindering synthesis and exploratory work. One participant noted, \textit{``Some of the databases were stood up in the 1980s''} (\blackcapsule{P04}), others observed that constrained availability \textit{``increases the risk of employees using unapproved AI tools \dots to do their job''} (\blackcapsule{P11}), prompting people to \textit{``look for their own solution''} when sanctioned tools could not accommodate available data (\blackcapsule{P27}). In these contexts, external AI tools were often perceived as more capable and immediately usable.

\subsubsection{Governance Gaps in Using Approved Technologies} \label{sec:governancegaps}
Participants identified ambiguities in governance and approval boundaries as a driver of shadow AI use, noting that employees were using \textit{``approved technology with data they have approved access to in ways that hadn't been assessed or approved,''} (\blackcapsule{P24}). This blurred the line between sanctioned and unsanctioned use, and was compounded by AI features being enabled within enterprise tools without formal risk assessment or governance approval. Several participants observed that \textit{``vendors were effectively turning [features] on''} (\blackcapsule{P11}), enabling employees to informally experiment and \textit{``see what it does for two days''} (\blackcapsule{P25}). As a result, security teams were placed on \textit{``the backfoot''} (\blackcapsule{P23}), with only \textit{``some level of awareness \dots but not a great detail”} of these capabilities (\blackcapsule{P13}), and AI use proceeding without prior assessment or integration into existing assurance processes. 

Governance gaps were not limited to general users. Participants described privileged technical roles extending informal practices, including administrators who \textit{``can install anything''} (\blackcapsule{P25}) and deploy tools \textit{``without seeking approval''} (\blackcapsule{P11}). Collectively, these accounts suggest that governance processes often focus on tools and nominal data access, while overlooking how embedded AI capabilities are used without assessment, and how privileged roles can bypass formal review. 

\subsubsection{Limited Digital and AI Literacy at Organisational Level} Several participants identified gaps in organisational AI understanding, distinct from individual interest or experimentation, as enabling informal and inconsistent use. While some employees actively explore AI, organisations reported gaps in shared knowledge about AI capabilities, risks, and appropriate use. Participants noted limited organisational expertise, stating they were \textit{``not experts in this thing''} and instead relied on \textit{external experts} (\blackcapsule{P01}) and \textit{``consultants coming to help us''} (\blackcapsule{P05}) to address what was described as \textit{``pretty immature''} organisational awareness of AI (\blackcapsule{P11}). 
Several participants noted that existing AI training and awareness initiatives did not address shadow AI use. One participant reported \textit{``No. Nothing, not yet''} when asked about training (\blackcapsule{P06}), while another noted that there was \textit{``no e-learning or anything formal \dots about Shadow AI''} (\blackcapsule{P19}). In the absence of formal guidance or training, staff had to rely on individual judgment to interpret acceptable use, increasing variation in practice and lowering barriers to shadow AI adoption.

\subsubsection{Misalignment Between Technology Adoption and Regulatory Cycles} Participants frequently described a mismatch between the rapid pace of AI innovation and slower organisational decision-making, budgeting, and approval processes in regulated environments. Long planning and regulatory cycles were seen to constrain timely AI adoption, with one interviewee noting, \textit{``If you're working in five-year planning cycles, you're not nimble enough to take advantage of an emerging technology''} (\blackcapsule{P06}). Another observed that four-year regulatory cycles could lead to missed opportunities if investment was not anticipated early (\blackcapsule{P10}). Cost constraints further delayed formal deployment, with participants noting that \textit{``one that is approved by the company \dots is going to cost money and that's the issue here''} (\blackcapsule{P27}). In this context, Shadow AI was framed as a stopgap, bridging immediate user demand and organisational readiness to fund sanctioned alternatives, with one participant noting that \textit{``once we get Copilot in, we'll get rid of people's subscriptions''}~(\blackcapsule{P25}). 

\begin{tcolorbox}[
  colback=gray!5,
  colframe=black!40,
  boxrule=0.6pt,
  rounded corners,
  boxsep=2pt,
  left=1.5mm,
  right=1.5mm,
  top=0.5mm,
  bottom=0.5mm
]
{\textbf{\ul{Key Insight:}} 
\textit{Shadow AI is primarily a response to productivity demands, cultural signals, governance lag, infrastructural gaps, and the pervasive availability of AI beyond organisational boundaries, rather than reckless or intentionally evasive behaviour.}}

\end{tcolorbox}

\subsection{Shadow AI Use Cases and Tools} \label{sec:usecases}

Participants described a consistent set of shadow AI use cases, centred primarily on text- and knowledge-intensive tasks rather than autonomous or system-integrated workflows. 

\subsubsection{Document Drafting and Content Generation} The most frequently reported uses involved drafting, editing, and summarising documents, including reports, emails, regulatory text, and presentation or training materials, primarily as cognitive support for routine work. As one participant explained, shadow AI was used to \textit{``look at what's out there publicly \dots and help put together a structure and things to consider''} as part of an initial whiteboarding process (\blackcapsule{P01}). Others noted similar applications for preparing PowerPoint decks (\blackcapsule{P14}) or responding to routine correspondence (\blackcapsule{P21}). In addition to ad hoc drafting, one interviewee noted that informal AI tools were being used to \textit{``generate training materials and things like that''} (\blackcapsule{P05}), indicating use for the development of training materials and workforce knowledge support. 

\subsubsection{Research and Information Seeking} A common use case involved information seeking and synthesis. Participants reported using informal AI tools to obtain background and contextual information on regulations, market conditions, and sector-specific technical material. One participant noted, \textit{``there's a lot of regulations, there's a lot of complexity \dots so they're using it for that'' } (\blackcapsule{P18}), while another described developing \textit{``a 10-page research report through Gemini \dots''} (\blackcapsule{P25}). In some cases, use extended to lightweight data tasks, such as uploading spreadsheets or reports for summaries or insights (\blackcapsule{P27}), though these activities were framed as informal and task-specific rather than systematic analytics workflows.

\subsubsection{Coding and Technical Assistance} Coding-related use cases were also reported, though less uniformly. Participants noted the use of AI tools to assist with code writing, debugging, or exploration, often in a supporting rather than autonomous role. One participant observed that \textit{``it's Claude for coding''} (\blackcapsule{P18}), while another described developers installing tools or spinning up models before seeking formal approval (\blackcapsule{P11}). However, such activity was typically characterised as episodic and exploratory rather than production-critical.

\subsubsection{Custom RAG and Internal Model Experiments} Beyond individual use of public AI tools, a small number of participants reported more advanced informal experimentation, including bespoke retrieval-augmented generation (RAG) systems. One participant noted that people were \textit{``spinning up their own RAG models and us not knowing about it until it gets to the point where they want to start sharing it''} (\blackcapsule{P05}). These efforts were described as bottom-up, locally scoped, and exploratory, and were typically discovered only once users sought broader organisational adoption. While indicating greater technical sophistication, they were not characterised as autonomous, agentic, or operating without human oversight.

\subsubsection{Tools and Platforms} Participants referenced several widely available, general-purpose generative AI systems, most commonly ChatGPT, followed by Claude for coding-related tasks, Gemini for search- and research-oriented use,
Canva for presentations, and NotebookLM for generating training materials. Other AI tools mentioned include Perplexity and Grok. At the same time, many organisations reported the parallel availability of sanctioned tools, most commonly Microsoft Copilot, which were positioned as safer substitutes but did not fully displace use of public AI tools (e.g., \blackcapsule{P07}, \blackcapsule{P10}, \blackcapsule{P18}).  

\begin{tcolorbox}[
  colback=gray!5,
  colframe=black!40,
  boxrule=0.6pt,
  rounded corners,
  boxsep=2pt,
  left=1.5mm,
  right=1.5mm,
  top=0.5mm,
  bottom=0.5mm
]
{\textbf{\ul{Key Insight:}} 
\textit{
Shadow AI is primarily used for productivity and cognitive support in knowledge work, rather than as an autonomous, system-integrated decision-making capability.}}\end{tcolorbox}

\subsection{Breaches and Severity} \label{sec:breaches}
Participants reported several incidents involving inappropriate use of generative AI tools. These were generally characterised as limited in scope and severity, detected through monitoring or data loss prevention (DLP) controls, and not associated with any malicious or intentional exfiltration.

\subsubsection{Sensitive Data Exposure} The most commonly reported incidents involved uploading organisational material into public generative AI tools. As discussed in Section~\ref{sec:usecases}, several participants reported sharing of engineering drawings, spreadsheets, regulatory submissions, and internal surveys via tools like ChatGPT. 
One organisation observed \textit{``a huge volume of documents being uploaded into ChatGPT''} based on log data (\blackcapsule{P12}), while another reported sharing of coding artefacts containing personal information, constituting a privacy breach (\blackcapsule{P18}). Others noted the disclosure of sensitive  (\blackcapsule{P14}) or \textit{``commercial in confidence''} information not intended for disclosure outside approved environments~(\blackcapsule{P17}). Some participants emphasised that the most sensitive systems were typically well protected through strong access controls. As one interviewee explained, \textit{``our sensitive systems are quite locked down and the people who are in charge of that are not many and well-trained''} (\blackcapsule{P18}). This suggests that while core systems were perceived as secure, sensitive data exposure still occurred at the enterprise level through informal AI interactions. 

\subsubsection{Perceived Severity and Impact} \label{sec:severity} Despite incidents, several participants downplayed their severity, emphasising they had not observed catastrophic consequences or national-level harm. While some acknowledged that \textit{``there have been some breaches''} involving material covered by the Privacy Act (\blackcapsule{P18}), others noted that \textit{``there was no customer data, so nothing that would be compliance related,''} even if it was information they \textit{``would rather never be in an outside tool''} (\blackcapsule{P17}). Several participants characterised the incidents as low impact, describing them as \textit{``real benign stuff''} (\blackcapsule{P08}) or stating \textit{``the risk is minimal''} (\blackcapsule{P01)}. Consistent with these assessments, organisations had \textit{``not seen enough risk yet to pull that trigger''} in terms of drastic enforcement actions or blanket bans (\blackcapsule{P18}). This illustrates how the lack of perceived immediate or severe harm shapes how organisations approach subsequent detection and response to shadow AI incidents.

\subsubsection{Detection and Organisational Response} Detection of shadow use was primarily attributed to technical monitoring and DLP tooling. Participants cited platforms such as Microsoft Defender (\blackcapsule{P11}), Purview (\blackcapsule{P26}), CrowdStrike (e.g., \blackcapsule{P09}, \blackcapsule{P10}), and Netskope (\blackcapsule{P13}) as enabling visibility into AI usage and data egress. 
DLP monitoring was flagged as a prominent detection mechanism, with one organisation reporting over 1,300 triggered events within a month (\blackcapsule{P14}); others stated that it is effective only \textit{``to some extent''} (\blackcapsule{P27)}. Responses varied across organisations, spanning selective blocking, provision of sanctioned alternatives, mandatory training, and ongoing monitoring with internal follow-up, as discussed in Section~\ref{sec:controls}. 

\subsubsection{Intent and Awareness} Across organisations, participants emphasised that breaches and near misses were driven primarily by lack of awareness and well-intentioned behaviour rather than malicious intent. Many described employees as unaware of what constituted sensitive information in the context of generative AI, or unclear about which tools were sanctioned. Several organisations framed such incidents as learning opportunities, reinforcing training and guidance rather than escalating enforcement (\blackcapsule{P09}, \blackcapsule{P15}, \blackcapsule{P18}). One participant summarised, \textit{``people were not trying to be malicious \dots they were just asking questions''} (\blackcapsule{P17}).

\begin{tcolorbox}[
  colback=gray!5,
  colframe=black!40,
  boxrule=0.6pt,
  rounded corners,
  boxsep=2pt,
  left=1.5mm,
  right=1.5mm,
  top=0.5mm,
  bottom=0.5mm
]

{\textbf{\ul{Key Insight:}} 
\textit{Shadow AI-related risk is characterised less by acute or catastrophic incidents than by largely benign, well-intentioned use patterns that, over time, incrementally increase exposure to more serious breaches.}}
\end{tcolorbox}


\subsection{Controls and Interventions} \label{sec:controls}

Participants reported a wide spectrum of controls and interventions used to manage shadow AI, reflecting differing organisational risk appetites, governance maturity, and innovation approaches. Rather than converging on a single approach, organisations were described as layering controls of varying strictness, combining technical restrictions with monitoring, education, and sanctioned alternatives.

\subsubsection{Prohibitive Controls} A small number of organisations adopted a prohibitive stance toward shadow AI, prioritising the outright blocking of non-approved tools. In some cases, access to public AI services was denied shortly after ChatGPT's release, even before sanctioned enterprise rollouts, with one participant noting such tools were \textit{``blocked pretty early on in the surge of ChatGPT's popularity,''} (\blackcapsule{P04}). Others implemented blanket bans only after deploying approved alternatives (e.g. \blackcapsule{P17}), while some selectively blocked specific tools (\blackcapsule{P03}, \blackcapsule{P10}, \blackcapsule{P15}). DeepSeek was the only system explicitly blocked by all organisations, following government advice\footnote{\url{https://www.protectivesecurity.gov.au/news/pspf-direction-update-deepseek-products-applications-and-web-services}}. 

\subsubsection{Detective Controls (Monitoring and Visibility)} More commonly, organisations relied on monitoring to gain visibility into shadow AI use rather than preventing it outright. Participants described widespread use of monitoring technologies to identify access to public AI services and detect data uploads to non-sanctioned tools, enabling organisations to observe usage patterns, flag potential data egress, and support follow-up actions. One participant described detailed reporting of \textit{``what people are doing, what they're using, where they're from''} (\blackcapsule{P22}). Monitoring was often framed as a deliberate alternative to strict blocking, aimed at preserving visibility and enabling engagement rather than enforcement. One participant explained, \textit{``our preference is not to block, it's to monitor, to understand, to educate, and respond''} (\blackcapsule{P24}), a view echoed across organisations (e.g., \blackcapsule{P09}, \blackcapsule{P15}, \blackcapsule{P22}). Monitoring was also positioned as a way to gauge demand and inform decisions on sanctioned frontier AI capabilities (e.g., \blackcapsule{P02}). 

\subsubsection{Behaviour-Shaping Controls} Several organisations described controls aimed at shaping user behaviour without outright prohibition, including software-based measures such as splash screens warning users not to share sensitive data (\blackcapsule{P18}) and nudge tools that guide users toward compliant behaviour (\blackcapsule{P21}). Others reported selectively disabling high-risk features, such as agent functionality or bring-your-own-license options, while leaving core capabilities accessible (e.g. \blackcapsule{P10}). 
Beyond technical controls, participants described the use of educational and conversational interventions to steer users toward sanctioned alternatives and reinforce appropriate use. These interactions were often enabled by monitoring but framed as dialogue rather than enforcement; for example, when staff expressed a preference for public tools such as ChatGPT or Claude, the response would be \textit{``did you know that in Copilot you can actually turn some of those models on \dots ?''} (\blackcapsule{P09}). Some organisations emphasised decentralising responsibility for these conversations, encouraging \textit{``[department] leaders to have conversations with their teams''} (\blackcapsule{P02}).

\subsubsection{Exception-Based Controls}

In some cases, organisations adopted exception-based approaches grounded in trust-based formal exception and risk-acceptance processes, as noted by one participant: \textit{``there needs to be an exception \dots people \dots accept the risk. ''} (\blackcapsule{P17}). Others adopted a path of \textit{``freedom within boundaries''} (\blackcapsule{P26}), a governance approach that allowed organisational units to exercise discretion in terms of experimenting and using AI tools. 

\subsubsection{Substitution-Based Controls} A dominant control strategy involved the provision of sanctioned AI tools as substitutes for shadow AI. Many organisations explicitly described rolling out internal or enterprise-approved solutions, most commonly Microsoft Copilot, to reduce reliance on public AI systems (e.g., \blackcapsule{P02}, \blackcapsule{P05}, \blackcapsule{P07} \blackcapsule{P10}, \blackcapsule{P18}). Participants emphasised that providing a safe, usable alternative was critical, noting that \textit{``if you don't give them the tools, they'll find a way to get the tools''} (\blackcapsule{P07}) and \textit{``by giving people a common tool, they'll use shadow AI less''} (\blackcapsule{P25}). Some organisations further enhanced these sanctioned tools by grounding them in internal data or offering secure sandpit environments for experimentation (\blackcapsule{P10}, \blackcapsule{P24}). In several cases, access to public AI tools was only restricted after sanctioned alternatives were made available, reflecting a deliberate substitution-first strategy (\blackcapsule{P17}, \blackcapsule{P24}).

\subsubsection{Cultural and Capability Controls} Across nearly all organisations, cultural and capability-based controls were central to managing shadow AI. Participants emphasised education, training, and awareness as foundational and often more effective than technical restrictions. Many organisations implemented mandatory training on generative AI use, data sensitivity, and individual accountability (e.g., \blackcapsule{P02}, \blackcapsule{P09}, \blackcapsule{P10}, \blackcapsule{P15}, \blackcapsule{P20}).  
Others reinforced these through policies, documentation, and internal communications, including AI-specific ethical principles and acceptable-use guidelines (\blackcapsule{P15}, \blackcapsule{P26}). Several participants explicitly framed their approach as avoiding overly heavy-handed controls that might appear dictatorial (\blackcapsule{P19}), \textit{``stifle innovation''} (\blackcapsule{P09}, \blackcapsule{P20}) or push use onto personal devices beyond organisational oversight (\blackcapsule{P13}, \blackcapsule{P24}, \blackcapsule{P27}).
However, while education and training initiatives strongly emphasised data handling and sensitivity, they rarely addressed shadow AI explicitly, leaving informal use largely outside existing control frameworks. 

\subsubsection{Limits of Blocking as a Control Strategy} Participants across organisations expressed scepticism that strict blocking could eliminate shadow AI use, instead viewing it as structurally persistent, driven by ease of access, personal devices, and embedded AI functionality within everyday tools. One participant noted, \textit{``it's really hard to control, particularly with people using their own devices''} (\blackcapsule{P05}); another remarked that restrictions are easily circumvented as users \textit{``can always take a screen capture or take a picture of their screen''} (\blackcapsule{P27}). Several participants cautioned that aggressive blocking may be counterproductive, pushing shadow AI use beyond organisational visibility rather than reducing it. As one participant noted, bans may shift use to personal devices, where \textit{``we can't see it at all and we can't have those conversations''} (\blackcapsule{P24}), while others warned that overly restrictive approaches risk undermining trust and stifling innovation (\blackcapsule{P20}).

Many participants framed informal AI use as inevitable, comparing attempts to ban generative AI to unrealistic restrictions on everyday tools such as pens, pencils or phones (e.g., \blackcapsule{P16}, \blackcapsule{P14}). Others described shadow AI as \textit{``just a new variant of the same problem''} organisations have long faced with shadow IT and consumer technologies (\blackcapsule{P08}). Against this backdrop, most organisations described a deliberate shift away from elimination-oriented strategies toward approaches that accept the persistence of shadow AI while seeking to manage its risks through visibility, dialogue, and layered controls. 

\begin{tcolorbox}[
  colback=gray!5,
  colframe=black!40,
  boxrule=0.6pt,
  rounded corners,
  boxsep=2pt,
  left=1.5mm,
  right=1.5mm,
  top=0.5mm,
  bottom=0.5mm
]
{\textbf{\ul{Key Insight:}} 
\textit{Organisational responses to shadow AI are moving towards layered controls that balance visibility, risk reduction, and productivity in the face of inevitable use.}}
\end{tcolorbox}

\subsection{Security, Compliance and Regulatory Risks}
\label{sec:nine_risks}
Several security-, compliance-, and regulatory-relevant risks emerge from observed shadow AI practices, based on participants' accounts and cross-interview analysis. While some risks were directly articulated, others were inferred from gaps in controls, governance, and regulatory obligations, together forming a set of interrelated shadow AI risk patterns. 

\subsubsection{Sensitive Data Exposure Beyond Organisational Assurance Boundaries} 

As discussed in Sections~\ref{sec:usecases} and~\ref{sec:breaches}, participants identified exposure of sensitive information as one of the most severe risks associated with shadow AI use, citing multiple instances of organisational information being shared with unapproved external AI systems. Although often framed as transient or informal, such interactions extended data handling beyond organisational assurance boundaries and enabled unauthorised disclosure.

At the same time, several participants downplayed the relevance of this risk within their own organisations. Some pointed to strong internal controls, asserting they were \textit{``not allowing any of our sensitive data to be touched''} (\blackcapsule{P02}) or had \textit{``really strong controls against shadow AI exposing sensitive [information]''} (\blackcapsule{P11}). Others discounted the risk due to the absence of visible harm or catastrophic outcomes (see Section~\ref{sec:severity}), or evaluated exposure primarily through a customer- or personal-data-centric lens, treating the absence of such data as indicative of lower risk. As one participant noted \textit{``It's not that the data leaking is not important, it's just not nearly as important for us''} (\blackcapsule{P17}). This framing downplays the security sensitivity of internal, operational, and contextual information, particularly in CI contexts. 

\begin{tcolorbox}[
  colback=gray!5,
  colframe=black!40,
  boxrule=0.6pt,
  rounded corners,
  boxsep=2pt,
  left=1.5mm,
  right=1.5mm,
  top=0.5mm,
  bottom=0.5mm
]
{\textbf{\ul{Key Insight:}} \textit{Sensitive data exposure emerges as one of the most severe risks associated with shadow AI use in CI contexts, with customer-data-centric framing exacerbating the problem by obscuring the security and national-interest significance of operationally sensitive information.}}
\end{tcolorbox}

\subsubsection{Insider-Based Exfiltration via Generative AI Interfaces}

One participant explicitly raised concerns that public generative AI interfaces could function as insider-driven data exfiltration pathways, enabling authorised users to move sensitive organisational information beyond approved security boundaries through routine interactions with external AI services. As the participant observed: \textit{``You can log onto ChatGPT, copy and paste some customer data across \dots and you'll have customers sitting in your history''} (\blackcapsule{P08}). While this framing was articulated explicitly in a single account, other interviews identified contextual conditions under which such a pathway could plausibly materialise. In particular, several participants reported the absence of formal monitoring or blocking of public generative AI tools, alongside a deliberate reliance on employee judgement and training rather than technical enforcement. Additionally, many organisations permitted access to public AI services on corporate devices as part of trust-based governance approaches, reducing friction for routine use while limiting the organisation's ability to prevent or detect intentional data exfiltration through these channels. 

\begin{tcolorbox}[
  colback=gray!5,
  colframe=black!40,
  boxrule=0.6pt,
  rounded corners,
  boxsep=2pt,
  left=1.5mm,
  right=1.5mm,
  top=0.5mm,
  bottom=0.5mm
]
{\textbf{\ul{Key Insight:}} \textit{Shadow AI constitutes a low-friction, high-plausibility channel for data egress that is structurally enabled by trust-based governance practices and difficult to detect or audit using existing security controls.}}
\end{tcolorbox}

\subsubsection{Loss of Assurance from Unassessed AI Use in Approved Systems}

Beyond external tools, interview analysis reveals a distinct risk from unassessed AI functionality embedded within sanctioned platforms. Participants described difficulties identifying when AI capabilities were introduced via scheduled or unscheduled system upgrades or new feature releases, noting that references to AI may be \textit{``embedded in small text''} (\blackcapsule{P10}). Others expressed similar frustration with vendor-driven change, observing that suppliers turn features on (\blackcapsule{P11}) or \textit{``just change stuff''} that may not meet security standards (\blackcapsule{P22}). Others expressed frustration that vendors \textit{``change the model whenever they want''} and have \textit{``no transparency of what model it is at any time''} (\blackcapsule{P17}). This complicates timely detection, preparation, and governance review. Combined with inquisitive employee behaviour where users informally experiment to \textit{``see what it does''} (\blackcapsule{P25}), this can directly undermine organisational assurance, auditability, and compliance.  This shifts the failure from isolated tool use to an organisational assurance gap, where AI‑mediated processing within approved systems expands effective access without visibility to monitoring or audit mechanisms. Consequently, organisations may struggle to evidence compliance, reconstruct AI involvement during incidents, or assess downstream risk, particularly in regulated or safety-critical contexts. This reflects a mismatch between static governance frameworks and dynamically evolving AI capabilities, rather than misuse of individual tools.

\begin{tcolorbox}[
  colback=gray!5,
  colframe=black!40,
  boxrule=0.6pt,
  rounded corners,
  boxsep=2pt,
  left=1.5mm,
  right=1.5mm,
  top=0.5mm,
  bottom=0.5mm
]
{\textbf{\ul{Key Insight:}} \textit{Unassessed AI capabilities embedded within approved systems create an assurance-bypass condition, allowing AI-mediated inference and aggregation to evolve beyond formal assessment, monitoring, and audit.}}
\end{tcolorbox}

\subsubsection{AI-Mediated Access Expansion from Unapproved Use Cases}

The interviews identified a shadow AI risk arising from unapproved AI use cases enabled by organisation-wide, AI-mediated inference and aggregation. Participants expressed concern that employees increasingly apply AI capabilities within approved tools to analyse data in ways that \textit{``hadn't been assessed or approved''} even when accessing data they are formally authorised to view (\blackcapsule{P24}). An illustrative example of this was AI-generated analysis revealing sensitive employee information such as \textit{``what someone else is getting paid or other private information''} without violating any underlying access-control (\blackcapsule{P01}). In the absence of explicit governance of these AI-mediated use cases (see Section~\ref{sec:governancegaps}), effective access expands through inference and synthesis rather than deliberate permission changes. 
Because such disclosures occur within sanctioned systems and workflows, they may evade access control, logging, and auditing mechanisms that assume a direct link between authorised data access and information disclosure. Over time, this undermines least privilege, data segregation, and role-based access control, creating persistent and hard-to-detect privacy, compliance, and insider-risk challenges that are difficult to detect or remediate post hoc.

\begin{tcolorbox}[
  colback=gray!5,
  colframe=black!40,
  boxrule=0.6pt,
  rounded corners,
  boxsep=2pt,
  left=1.5mm,
  right=1.5mm,
  top=0.5mm,
  bottom=0.5mm
]
{\textbf{\ul{Key Insight:}} \textit{Unapproved AI use cases within sanctioned systems enable AI-mediated inference and aggregation that expands effective access to sensitive information beyond established access-control assumptions.}}
\end{tcolorbox}

\subsubsection{Decision and Epistemic Integrity Failure from Unvetted Model Outputs}

Several participants raised concerns regarding decision-integrity failures arising from informal reliance on unvetted generative AI outputs. Several noted that while AI-generated drafts, analyses, or code are increasingly incorporated into work products, users may not consistently verify or critically assess these outputs. One participant asked, \textit{``Do they fully review it? \dots they should''} (\blackcapsule{P25}) because \textit{``we don't know the quality of the outcome''}(\blackcapsule{P10}). This risk is amplified by limited organisational understanding of model behaviour and persistent misconceptions about reliability and determinism. Participants emphasised that generative models are non-deterministic and frequently incorrect, noting that \textit{``It does hallucinate \dots it's all you''}~(\blackcapsule{P18}). Despite this, others observed that people may still accept outputs uncritically because \textit{``that's what Copilot or ChatGPT said''}~(\blackcapsule{P14}). 

These accounts describe a decision-integrity failure mode in which automation bias and unvetted model performance allow incorrect or fabricated information to propagate into operational artefacts, codebases, or decisions. Because AI use is often informal and unlogged, accountability becomes diffuse and error attribution difficult. Over time, this weakens organisational sense-making and creates hard-to-audit failures that may only emerge once downstream harms have occurred.

\begin{tcolorbox}[
  colback=gray!5,
  colframe=black!40,
  boxrule=0.6pt,
  rounded corners,
  boxsep=2pt,
  left=1.5mm,
  right=1.5mm,
  top=0.5mm,
  bottom=0.5mm
]
{\textbf{\ul{Key Insight:}} \textit{Informal reliance on unvetted AI outputs within shadow AI practices undermines decision and epistemic integrity, enabling incorrect, biased, or fabricated information to propagate into organisational artefacts and decisions without clear accountability or auditability.}}
\end{tcolorbox}

\subsubsection{Opaque Third-Party and Supply Chain Dependencies} 

Concerns regarding third-party and supply-chain dependencies featured prominently across interviews, though views varied. Participants highlighted limited visibility into upstream system changes and downstream data use, questioning whether contractual safeguards meaningfully constrained secondary data use or model improvement activities. One participant noted uncertainty about \textit{``how our data is being used to further the interests of vendors''} (\blackcapsule{P21}), while others raised concerns about vendors introducing changes with little notice, stating they \textit{``just change stuff''} without meeting internal security or governance standards (\blackcapsule{P22}). Related concerns extended to broader supply-chain assurance, \textit{``what internal controls do they have themselves?''} (\blackcapsule{P02}), as well as implications of foreign control or foreign ownership 
(\blackcapsule{P09}). 
Some participants expressed strong confidence in major AI vendors, describing them as \textit{``big trusted vendors''} (\blackcapsule{P25}) and \textit{``international juggernauts whose reputation would be critical''} (\blackcapsule{P08}) and therefore unlikely to mishandle data. For them, shadow AI-related third-party risks were considered manageable within existing procurement and vendor assurance processes.

Collectively, these accounts suggest that third-party shadow AI risk is often viewed through existing trust and procurement lenses rather than as a distinct security failure mode. This framing obscures how routine shadow AI use can introduce dependencies on external providers, shifting risk beyond organisational control. As a result, organisations may unknowingly extend their trust boundaries and attack surface while lacking the visibility and assurance needed to assess compliance, trace data flows, or respond to incidents.

\begin{tcolorbox}[
  colback=gray!5,
  colframe=black!40,
  boxrule=0.6pt,
  rounded corners,
  boxsep=2pt,
  left=1.5mm,
  right=1.5mm,
  top=0.5mm,
  bottom=0.5mm
]
{\textbf{\ul{Key Insight:}} \textit{Third-party shadow AI silently shifts organisational trust boundaries to external vendors by introducing opaque, vendor-controlled dependencies that undermine visibility, governance, and assurance.}}
\end{tcolorbox}

\subsubsection{Loss of Organisational Observability and Security Situational Awareness} 

Participants repeatedly described shadow AI use occurring outside formal monitoring and control regimes, particularly where access to public AI tools was deliberately permitted or shifted to personal devices. Often justified as a pragmatic trade-off to preserve productivity or avoid pushing use underground, these practices reduce organisational visibility into how AI is used, what data it processes, and how outputs influence work. From a security perspective, this constitutes a loss of organisational observability. When AI interactions are unlogged or occur beyond enterprise boundaries,  situational awareness of AI-mediated data flows and decision influence is diminished, weakening detection, impairing incident response, and limiting the organisation's capacity to assess the scope or impact of shadow AI use when issues arise. Over time, diminished observability undermines security governance by decoupling organisational risk awareness from actual AI use in practice.

\begin{tcolorbox}[
  colback=gray!5,
  colframe=black!40,
  boxrule=0.6pt,
  rounded corners,
  boxsep=2pt,
  left=1.5mm,
  right=1.5mm,
  top=0.5mm,
  bottom=0.5mm
]
{\textbf{\ul{Key Insight:}} \textit{Shadow AI erodes organisational observability and security situational awareness, obscuring AI‑mediated data flows and decision influence while impairing detection, incident response, and governance.}}
\end{tcolorbox}

\subsubsection{Regulatory and Compliance Risk from Evidentiary Gaps}

Several participants described shadow AI interactions as informal, conversational, and transient, with limited documentation or an audit trail. While often viewed as benign in the absence of immediate harm, such practices introduce regulatory and compliance risk in regulated CI environments. Because informal AI influence on documents, analyses, or operational decisions is frequently undocumented or untracked, organisations may be unable to reconstruct how outputs were produced, what data was used, or where accountability lies. In contexts where regulatory frameworks emphasise traceability and post-incident evidentiary requirements, these gaps complicate audits, incident reporting, and assurance activities, even in the absence of an explicit policy breach. As a result, organisations may remain nominally compliant while lacking the evidentiary basis to demonstrate compliance when scrutiny is applied.

\begin{tcolorbox}[
  colback=gray!5,
  colframe=black!40,
  boxrule=0.6pt,
  rounded corners,
  boxsep=2pt,
  left=1.5mm,
  right=1.5mm,
  top=0.5mm,
  bottom=0.5mm
]
{\textbf{\ul{Key Insight:}} \textit{Shadow AI use without documentation or audit trails creates evidentiary gaps that undermine regulatory and compliance obligations by limiting organisations' ability to reconstruct AI involvement, demonstrate accountability, or support audits and post-incident investigations.}}
\end{tcolorbox}

\subsubsection{Risk Normalisation and Governance Drift}

Across organisations, participants commonly emphasised that shadow AI incidents had not resulted in visible harm, catastrophic outcomes, or national-level consequences. This absence of immediate impact was frequently cited to justify tolerance of informal use, preference for education over enforcement, or delayed intervention. Over time, this pattern contributes to risk normalisation, in which repeated low-severity boundary crossings become accepted as routine practice. As shadow AI use becomes embedded in everyday workflows, organisational expectations around acceptable use shift without formal governance updates or explicit risk acceptance decisions, allowing structurally enabled risks to persist and compound, reducing organisational sensitivity to early warning signals and delaying corrective action until more consequential harms emerge.

\begin{tcolorbox}[
  colback=gray!5,
  colframe=black!40,
  boxrule=0.6pt,
  rounded corners,
  boxsep=2pt,
  left=1.5mm,
  right=1.5mm,
  top=0.5mm,
  bottom=0.5mm
]
{\textbf{\ul{Key Insight:}} \textit{The absence of visible harm from repeated shadow AI use normalises informal boundary crossings, driving governance drift that allows risks to persist, compound, and evade timely corrective action.}}
\end{tcolorbox}


These findings collectively indicate that shadow AI represents a systemic condition characterised by recurring use patterns, control gaps, and governance misalignment. 
\section{Shadow AI Threat Model} \label{sec:dicussion}

Drawing on the empirical findings in Section~\ref{sec:results}, we derive a threat model that frames shadow AI as a security-relevant socio-technical condition in which generative-AI capabilities enter organisational workflows outside established assurance boundaries. To make this condition analytically tractable, we abstract shadow AI around assurance boundaries and recurring exposure pathways observed in practice. 

\subsection{System Model and Trust Boundaries}

The model considers CI environments where regulatory compliance, auditability, and operational integrity are central. Assurance requires not only preventing unauthorised disclosure, but also maintaining bounded systems, visible risk-bearing activities, demonstrable controls, and attributable accountability. We define the \textit{assurance boundary} as the set of organisational systems, governance processes, and technical controls (e.g., logging, monitoring, DLP, access control) that ensure data flows and decision-making processes remain observable and auditable. Shadow AI arises when AI-mediated data flows, transformations, or inference occur partially or wholly outside this boundary.

\subsection{Scope and Assumptions}

The model considers informal, interactive human-in-the-loop use of generative AI, excluding autonomous agentic systems deployed in production or control workflows. We assume an \textit{honest-but-curious} or \textit{productivity-optimising} internal actor, where use is widespread, persistent, and primarily efficiency-driven, and that existing technical controls were not designed for this form of AI-mediated use. 

\subsection{Assets at Risk}

Under this model, assets at risk extend beyond personal or customer data to include \textit{operational and organisationally sensitive information} (e.g. internal documents, engineering artefacts, analyses, code,  regulatory drafts); \textit{decision and knowledge integrity} (correctness, provenance, and accountability of AI-influenced outputs incorporated into operational or policy artefacts); \textit{assurance and auditability} (the ability to demonstrate regulatory compliance and evidence due diligence); and \textit{trust boundaries and control assumptions} (where data is processed, how inference occurs, and which systems mediate aggregation and transformation). In CI environments, these assets matter because secure and compliant operation depends on demonstrable oversight and attributable decision-making, not solely on confidentiality. 

\subsection{Adversary Model and Capabilities}

Risk originates from legitimate organisational actors rather than external adversaries, including general staff leveraging AI for routine tasks, privileged users extending tools' functionality without formal review, and trusted third-party vendors whose evolving models operate outside organisational control. The dominant risk is \textit{capability amplification} without commensurate governance or visibility, reframing the threat from malicious intent to unmanaged exposure pathways. Actors are assumed to be \textit{honest-but-curious} or \textit{productivity-optimising}. They can (i) submit organisational data to external or embedded AI systems, (ii) use AI to transform, summarise, or infer over this data, (iii) introduce or enable new AI capabilities without reassessment, and (iv) integrate AI-generated outputs into workflows without provenance tracking or audit visibility.

\subsection{Exposure Pathways and Attack Surface}

We identify three pathways through which shadow AI crosses organisational boundaries, driving assurance erosion. 

\subsubsection{External Public AI Use (Boundary Bypass)} External use of consumer-facing services for coding, drafting, or analysis occurs outside approved organisational controls, bypassing the assurance boundary and resulting in unmonitored data processing and loss of egress visibility. Empirically, such use was described as low-friction and routine, with organisational awareness often limited to its use rather than its purpose.

\subsubsection{Unassessed AI Capability (Capability Expansion and Usecase Drift)} New AI functionality in sanctioned enterprise platforms may be introduced without reassessment of security, governance, or assurance implications, expanding what can be inferred or aggregated and creating assurance blind spots within approved systems. Participants noted that such changes often arise from vendor updates or feature activations, bypassing established review processes and shifting system capabilities without explicit policy violations. Similar blind spots arise from unauthorised or unanticipated use cases, where users repurpose approved tools beyond their assessed scope. While not necessarily malicious, such uses may involve sensitive data combinations, novel inference pathways, or unassessed decision-support roles. Together, vendor-driven capability expansion and user-driven use-case drift widen the gap between assessed and actual system behaviour, undermining governance and assurance controls.  

\subsubsection{Privileged/Local Extension (Governance Circumvention)} Technically capable users may enable AI features, deploy local models, or develop local AI systems (e.g., RAG-style prototypes) outside formal governance processes. This often involves direct installation, external APIs, or integration with local data, bypassing established data-handling reviews, approved workflows, and logging mechanisms. As a result, such systems may access sensitive data without oversight, operate without audit trails, and circumvent access controls, eroding least-privilege enforcement and reducing organisational observability.

\subsection{Threat Actors and Risk Origins} 

Risk originates from legitimate organisational actors rather than external adversaries, including general staff leveraging AI for routine tasks, privileged users extending tools functionality without formal review, and third-party vendors whose evolving models operate outside local organisational control. The dominant risk is \textit{capability amplification} without commensurate governance or visibility, reframing the risk from malicious intent to unmanaged exposure pathways. 

\subsection{Security Properties and Threat Characteristics}

Across these exposure pathways, shadow AI undermines key security and assurance properties, including confidentiality (uncontrolled data exposure), integrity (non-deterministic or unverified transformations), accountability (loss of provenance), auditability (inability to reconstruct decisions), and observability (loss of visibility into data flows).

The resulting risk exhibits several defining properties. It is \textit{diffuse}, embedded in everyday workflows, and \textit{opaque}, remaining partially or wholly unobservable by existing monitoring and DLP mechanisms. This opacity is compounded by the \textit{non-deterministic} nature of generative models, where stochastic outputs hinder detection and forensic reconstruction. The risk is inherently \textit{socio-technical}, arising from interactions between human behaviour, tool affordances, and organisational governance structures. It is also \textit{incremental and cumulative}, weakening security and assurance over time. Finally, shadow AI risk is \textit{self-reinforcing}, as repeated low-severity boundary crossings normalise unsafe practices and reduce the perceived urgency of intervention. Participants frequently reported that the absence of serious incidents diminished concern, contributing to the gradual erosion of governance effectiveness.

\subsection{Threat Model Implications}

This model characterises shadow AI as a shift from adversarial compromise to assurance erosion. Rather than a single point of failure, it represents a structural misalignment where operational demands and AI capabilities outpace governance systems. Traditional block-or-permit security approaches are insufficient for this threat class, which requires frameworks capable of addressing decentralised, AI-mediated inference.

\section{Related Work} \label{sec:relwork}

\subsection{Conceptualising Shadow AI}

Across academic, industry, and policy discourse, shadow AI is defined as the informal, unsanctioned, or unmonitored use of AI  that bypasses formal approval, central oversight, and established governance controls~\cite{ kshetri2025transforming,sikos2025securing,puthal2025shadow}. Much of the literature frames shadow AI as an extension of shadow IT, enabled by the accessibility, low cost, and ease of use of contemporary AI tools, notably generative models and agentic systems~\cite{cerchione2026artificial,puthal2025shadows, ross2025shadow,chin2025conflicting, wendt2025}. In this context, some authors describe this phenomenon as \textit{Bring Your Own AI} (BYOAI)~\cite{van2024bring, arnaout2025you} 
However, several studies argue that shadow AI differs fundamentally from traditional shadow IT because it operates primarily at the interaction layer, enabling immediate, difficult‑to‑reverse security impacts without deployable artefacts. Because its use requires minimal technical integration and is enacted through user prompts or data inputs~\cite{sikos2025securing, oktenli2025strategic,ogungbemi2025shadow}, shadow AI disrupts governance and assurance approaches that rely on system visibility, asset inventories, and deterministic system behaviour~\cite{Silic2025FromShadow,ross2025shadow, ogungbemi2025shadow, huang2024build}.

\subsection{Drivers of Shadow AI} \label{sec:drivers}

Existing literature identifies a consistent set of drivers for shadow AI adoption, led by productivity pressures and perceived efficiency gains, particularly in contexts where sanctioned tools lag behind user needs or are constrained by slow procurement and approval~\cite{van2024bring,puthal2025shadows, wendt2025, te2026organizational, hawkins2026embracing}. The rapid proliferation and accessibility of generative AI tools, particularly LLMs and associated APIs, further lower barriers to informal adoption by enabling use without technical integration or formal authorisation~\cite{matsumoto2025a2a, puthal2025shadows}. 
Organisational and cultural factors also play a significant role, including fragmented governance structures, unclear AI policies, limited AI literacy, and the absence of sanctioned alternatives or safe experimentation environments~\cite{nrg2024outoftheshadows}. Where formal controls are perceived as misaligned with operational realities, employees often turn to external AI tools out of necessity rather than deliberate non-compliance~\cite{ross2025shadow}. Incentive structures that prioritise efficiency and innovation without commensurate governance safeguards~\cite{ross2025shadow, wendt2025}, as well as overly restrictive policies, can further catalyse shadow AI use~\cite{van2024bring,bonnin2026challenges}. 
Collectively, the literature frames shadow AI as a structural response to misalignment between escalating work demands and governance frameworks that cannot keep pace with modern AI capabilities.

\subsection{Risk Narratives and Governance Challenges}

Shadow AI introduces several risks, including inadvertent data leakage, loss of intellectual property, regulatory non-compliance, and expansion of the organisational attack surface~\cite{matsumoto2025a2a, puthal2025shadows, sikos2025securing, kwan2024navigating, puthal2025shadow}. Generative models amplify these risks due to their opaque operation, potential retention or reuse of input data, and propensity to generate plausible but incorrect outputs (hallucinations)~\cite{sikos2025securing,ross2025shadow, singh2025operationalizing,sikos2025generative}. 
Shadow AI is often described as a governance blind spot, marked by limited organisational visibility into where, how, and by whom AI tools are used until risks emerge or incidents occur~\cite{cerchione2026artificial, sikos2025securing, ross2025shadow, ogungbemi2025shadow, alhashimi2025exploring}. This lack of visibility undermines traditional accountability and assurance mechanisms, particularly in regulated environments where liability, auditability, and explainability are essential~\cite{puthal2025shadows, ogungbemi2025shadow, whitelaw2025rethinking, achanta2025data}. From a security perspective, shadow AI is increasingly seen as insider risk driven by optimisation-oriented workarounds rather than malicious intent~\cite{oktenli2025strategic, whitelaw2025rethinking}, with recent literature conceptualising shadow AI as a structural governance failure, and arguing that restrictive governance regimes can incentivise informal AI use and exacerbate risk~\cite{van2024bring, Mansner2025ShadowAI, wendt2025, ogungbemi2025shadow, hawkins2026embracing}. Seen through this lens, shadow AI signals persistent misalignment between formal governance structures and the realities of contemporary work practices.

\subsection{Mitigation Strategies and Proposed Interventions}

The literature proposes a range of technical, organisational, and policy-level responses to shadow AI. Common technical measures include AI asset discovery, centralised model registries, access control, DLP, and AI security posture management tools aimed at detecting and monitoring unauthorised AI usage~\cite{matsumoto2025a2a, sikos2025securing, huang2024build, puthal2025shadow, huang2025commercial, syed2024emerging}. Several sources advocate for inventories or ``AI bills of materials'' as foundational governance mechanisms~\cite{huang2025commercial, sayles2024, harjika2026strategic}. 
Organisational and behavioural interventions are also prominent, including AI literacy and awareness training, acceptable use policies, non-punitive disclosure mechanisms (e.g., shadow AI ``amnesty'' programs), and sanctioned alternatives or sandboxed environments~\cite{puthal2025shadows, sikos2025securing,wendt2025, te2026organizational, bonnin2026challenges, jackson2025aimaturity}. Notably, outright bans are explicitly criticised as counter-productive, as they tend to drive AI use underground~\cite{van2024bring, hawkins2026embracing, bonnin2026challenges, clarke2023ai}. 
However, much of the mitigation literature is prescriptive and normative, often assuming that shadow AI activity is already known, discoverable, or readily classifiable~\cite{huang2024build, huang2025commercial}. Moreover, only a small subset of studies draw on empirical organisational data to examine how mitigations emerge, reshape, or legitimise informal AI practices in real-world settings~\cite{te2026organizational}.

\subsection{Industry Insights on Shadow AI}

Industry surveys show that shadow AI has moved from a peripheral issue to a \textit{persistent security concern}, driven by individual adoption that outpaces organisational governance. Large-scale global studies report that most employees use generative AI for work purposes, often via publicly available or personally provisioned tools rather than employer-sanctioned systems, and frequently without explicit organisational approval or oversight~\cite{microsoftworktrend2025, gillespie2025trust}. Australian evidence mirrors these findings, showing widespread shadow AI use, routine sharing of sensitive and highly sensitive data with public AI tools, elevated risk in regulated sectors~\cite{josys2025ShadowAI}, and continued recognition of shadow AI as a major security blind spot~\cite{okta2026SecurityNews}.

Security-focused analyses show that \textit{shadow AI usage is widespread and persistent}. Unsanctioned AI applications often persist within enterprise environments for extended periods, account for an excessive share of observed shadow AI activity, and are increasingly linked to incidents involving sensitive data exposure and intellectual property loss~\cite{Silic2025FromShadow, recoshadowAI2025, ibmcostdatabreach2025}.

Reported shadow-AI use cases are largely productivity- and knowledge-focused, including document drafting and summarisation, software development and code review, data analysis, report and presentation preparation, customer communication, and operational decision support~\cite{microsoftworktrend2025, snowflake2026ROI, salesforcemuleSoft2026}. These activities are commonly enabled through consumer-grade tools, personal accounts~\cite{Mansner2025ShadowAI}, or embedded AI features within enterprise platforms, reducing organisational visibility, auditability, and effective governance. While the literature presents shadow AI as a normalised global phenomenon, it offers limited insight into how it is perceived, managed, and governed in practice, particularly in highly regulated sectors.

\subsection{Gaps in Existing Work}

Despite growing scholarly and industry attention, significant gaps persist in research on shadow AI. 
First, there is a lack of fine-grained empirical studies examining shadow AI as an organisational phenomenon. Most existing work relies on surveys, industry reports, or illustrative incidents, offering limited insight into how shadow AI manifests in practice. 
%
Second, academic research has largely overlooked shadow AI in CI environments, focusing predominantly on general organisational settings, with only cursory reference to safety-critical contexts. 
%
Third, existing work primarily enumerates technical or regulatory risks (e.g., data leakage, audit failures) while overlooking how these are weighed against operational pressures (e.g., service continuity, operational efficiency). As a result, shadow AI remains under-examined as a situated organisational practice shaped by multiple interacting factors, including operational demand, organisational culture, governance constraints, and security capability gaps. 
%
%
Fourth, there is limited insight into how organisational decision-makers perceive shadow AI in practice. In particular, little is known about how judgements about risk, necessity, and legitimacy shape governance and mitigation decisions as informal AI use becomes embedded in organisational workflows.
%
Finally, proposed mitigation strategies are largely prescriptive and weakly grounded in organisational reality.
Consequently, there is little understanding of how governance, assurance, and detection measures function, or break down, in practice. 
Collectively, these gaps motivate an empirical examination of shadow AI as a security-relevant organisational phenomenon within CI environments. By focusing on senior decision-makers, this study shows how shadow AI emerges from the persistent misalignment between operational demand, capability gaps, and governance enforcement under regulatory constraints.

\section{Conclusion} \label{sec:conclusion}

This paper presents the first empirical study of shadow AI in CI environments, showing it is widespread, embedded in routine knowledge work, and predominantly human-in-the-loop, with little evidence of autonomous deployment. Rather than being driven by malicious actors, shadow AI arises from productivity-focused practices, enabled by governance lag and the widespread accessibility of frontier AI. It introduces risks beyond data exposure, including failures in assurance, observability, least-privilege enforcement, and decision traceability, emerging through three mechanisms that normalise shadow AI while progressively eroding CI security assurance: \textit{boundary bypass}, where AI-mediated processing occurs outside approved controls; \textit{unassessed capability expansion}, where AI enables new inference and aggregation beyond established governance assumptions; and \textit{loss of observability (via governance circumvention)}, where organisations cannot reliably monitor or reconstruct AI-mediated activity. Shadow AI challenges core assumptions in CI assurance frameworks, particularly system boundedness, visibility, and demonstrable control. As its risks emerge gradually through routine practice rather than discrete incidents, prevention-driven approaches are insufficient. Instead, shadow AI should be understood as systemic assurance erosion, requiring governance approaches that address AI-mediated inference, decentralised use, and evolving capabilities. More broadly, this reflects a structural misalignment between operational demands, AI capability, and governance systems, rendering one-size-fits-all controls ineffective and necessitating tailored, pathway-aligned responses.

\section{Acknowledgements}

This work was supported by the UK AI Security Institute (AISI) via the AISI Challenge Fund (Grant Title: Frontier AI in Critical Infrastructure: Empirical Insights for Building Societal Resilience).


\section{Ethical Considerations}



This study involves human participants and examines the use of frontier AI in CI systems, raising ethical concerns around confidentiality, sensitive operational information, and security risks.

\noindent \textbf{Participant protection and data handling.} 
The study received approval from an institutional human research ethics committee (Approval \#: H-168/25) and adhered to established ethical guidelines. Participation was voluntary, with informed consent obtained from all participants, who were free to withdraw at any time without consequence. All data were de-identified, and sensitive details removed or generalised to prevent re-identification.\\
\textbf{Sensitivity of CI contexts.}
Given that participants discussed cybersecurity and system resilience, their accounts may reference vulnerabilities or coordination gaps. To minimise risk, findings are reported at an aggregated level, and quotations are selectively included and sufficiently generalised to prevent identification of sensitive information. \\
\noindent \textbf{Risk-benefit considerations.} 
While the study may reveal vulnerabilities and threats, risks are mitigated through anonymisation and abstraction. The primary benefit lies in improved understanding of system-level resilience, thereby improving governance and security practices.\\
\noindent \textbf{Dual-use implications.} 
Although findings highlight some gaps, the analysis remains conceptual and does not provide actionable details that could be exploited. The intent is to inform defensive and resilience-enhancing strategies rather than enable misuse. \\
\noindent \textbf{Data sharing.} 
Raw interview transcripts are not publicly shared due to re-identification risks. Instead, anonymised supporting materials (e.g., interview protocol, coding framework, and audit trail) are provided to support transparency. \\
\noindent \textbf{Researcher responsibility.} 
Findings are presented in a balanced manner, focussing on structural and system-level insights rather than organisational attribution.\\
Overall, the study seeks to balance transparency with the obligation to protect participants and minimise harm within a sensitive domain.


\bibliographystyle{IEEEtran}
\bibliography{bib}


\onecolumn
\appendix

\section*{Appendix A: Qualitative Codebook and Supporting Evidence}


This appendix presents the qualitative codebook used to organise and interpret interview data throughout the empirical analysis. Table~\ref{tab:shadow-ai-codebook} summarises the main themes, codes, operational interpretations, and illustrative quotations that underpin the findings presented in Section~\ref{sec:results}. Its purpose is to provide transparency into the analytic structure of the thematic analysis while preserving participant and organisational confidentiality.

The codebook is organised around key empirical themes reported in the paper: prevalence of shadow AI, drivers of shadow AI, use cases and tools, breaches and perceived severity, controls and interventions, and security, compliance, and regulatory risks. Each code reflects a recurring pattern identified across interviews, rather than a statistically estimated category. The illustrative quotations are included to support the interpretation of each code.

We do not report frequencies or percentages for individual themes. This reflects the qualitative, interpretive nature of the study, where the aim is to capture meaning, variation, and underlying patterns rather than quantify their occurrence. Reporting counts could imply a level of statistical generalisability that is not supported by the study design. Additionally, given the sensitivity of CI and cybersecurity contexts, even aggregate frequencies may risk revealing patterns about organisational practices or vulnerabilities. Instead, the analysis emphasises the salience, consistency, and explanatory value of themes across the dataset.


\newlength{\cbcolTheme}
\newlength{\cbcolA}
\newlength{\cbcolB}
\newlength{\cbcolC}
\newlength{\cbcolD}

\setlength{\cbcolTheme}{0.03\linewidth}
\setlength{\cbcolA}{0.18\linewidth}
\setlength{\cbcolB}{0.40\linewidth}
\setlength{\cbcolD}{\dimexpr\linewidth-\cbcolTheme-\cbcolA-\cbcolB-\cbcolC-4\tabcolsep\relax}

\footnotesize
\setlength{\LTleft}{0pt}
\setlength{\LTright}{0pt}
\setlength{\tabcolsep}{2pt}

\begin{longtable}{@{}L{\cbcolTheme}|L{\cbcolA}L{\cbcolB}L{\cbcolD}@{}}
\caption{Shadow AI in CI: qualitative codebook and supporting evidence.}\label{tab:shadow-ai-codebook}\\
\toprule
\multicolumn{2}{@{}l}{\textbf{Theme and Code}} & \textbf{Definition / Interpretation} 
& \textbf{Illustrative Quotation} \\
\midrule
\endfirsthead

\toprule
\multicolumn{2}{@{}l}{\textbf{Theme and Code}} & \textbf{Definition / interpretation} 
& \textbf{Illustrative quotation} \\
\midrule
\endhead

\bottomrule
\endfoot

\cbtheme{12}{\textsc{\textbf{Prevalence of Shadow AI}}}
& Overall prevalence
& Participants consistently described shadow AI as widespread, expected, and normalised across organisations.
& ``anyone that says no to that is probably dreaming'' \blackcapsule{P18} \\
\cline{2-4}

& Organisational spread
& Shadow AI use was described as diffused across teams, roles, and work settings rather than confined to technical specialists.
& ``It's across the organisation'' \blackcapsule{P18}\\
\cline{2-4}

& Persistence beyond sanctioned tools
& Informal AI use emerged before governance and continued even after sanctioned alternatives became available.
& ``By the time we put those policies in place, people were already using ChatGPT \ldots{} and the inertia of that has continued'' \blackcapsule{P05} \\
\cline{2-4}

& Absence of shadow agentic  AI
& Participants reported widespread interactive, human-in-the-loop use of generative AI, but not covert agentic or autonomous workflows.
& ``It's just chat, no one is building giant LLMs, agentic agents, or anything like that in the dark, in the shadows'' \blackcapsule{P08} \\
\midrule

\cbtheme{19}{\parbox{2.5cm}{\centering\textsc{\textbf{Drivers of Shadow AI}}}}
& Productivity pressures and recognised efficiency gains
& Shadow AI was framed as a pragmatic response to delivery pressure, routine drafting work, and recognised efficiency gains.
& ``a good initial whiteboarding process'' \blackcapsule{P01}\\
\cline{2-4}

& Ease of access and low barriers to adoption
& Public AI tools were perceived as easy to access across devices and personal accounts, reducing friction for informal uptake.
& ``Everybody has Gemini on their phone'' \blackcapsule{P27} \\
\cline{2-4}

& Executive signalling and permission to experiment
& Leadership openness to AI experimentation lowered social and psychological barriers to informal AI use.
& ``People were not trying to be malicious \ldots{} they were just asking questions \ldots{}'' \blackcapsule{P17}\\
\cline{2-4}

& Infrastructural limitations and fragmented data
& Legacy systems, fragmented data, and limited internal tooling encouraged users to seek external AI-enabled workarounds.
& ``Some of the databases were stood up in the 1980s'' \blackcapsule{P04} \\
\cline{2-4}

& Governance gaps in using approved technologies
& AI capabilities embedded in approved tools, together with privileged bypasses and incomplete oversight, created assurance blind spots.
& ``using approved technology with data they have approved access to in ways that hadn't been assessed or approved'' \blackcapsule{P24} \\
\cline{2-4}

& Limited digital and AI literacy at organisational level
& Participants described immature organisational understanding of AI capabilities, risks, and acceptable use, with little formal guidance on shadow AI.
& ``the awareness of what AI is [is] pretty immature'' \blackcapsule{P11}\\
\cline{2-4}

& Governance lag and adoption-cycle misalignment
& Long planning cycles, regulatory timing, and cost constraints delayed sanctioned AI deployment and sustained informal use.
& ``If you're working in five-year planning cycles, you're not nimble enough to take advantage of an emerging technology'' \blackcapsule{P06} \\
\midrule

\cbtheme{13}{\textsc{\textbf{Shadow AI Use Cases \& Tools}}}
& Document drafting and content generation
& Shadow AI was most commonly used to draft, edit, structure, and generate routine text-based work products.
& ``look at what's out there publicly \ldots{} and help put together a structure and things to consider'' \blackcapsule{P01} \\
\cline{2-4}

& Research and information seeking
& Participants used generative AI for information gathering, synthesis, and lightweight analysis in complex regulatory and technical environments.
& ``there's a lot of regulations, there's a lot of complexity \ldots{} so they're using it for that'' \blackcapsule{P18} \\
\cline{2-4}

& Coding and technical assistance
& AI tools were also used episodically to support coding, debugging, and technical exploration.
& ``it's Claude for coding'' \blackcapsule{P18}\\
\cline{2-4}

& Custom RAG and internal model experiments
& A smaller set of technically capable users conducted bottom-up experimentation with internal RAG-style systems outside formal oversight.
& ``spinning up their own RAG models and us not knowing about it until it gets to the point where they want to start sharing it with the rest of the organisation'' \blackcapsule{P05} \\
\cline{2-4}

& Tools and Platforms
& Participants referenced several widely available, general-purpose generative AI systems.
& ``it's Claude for coding'' \blackcapsule{P18} \\

\newpage
\cbtheme{11}{\parbox{2.5cm}{\centering\textsc{\textbf{Breaches \& Severity}}}}& Sensitive data exposure
& The most commonly reported incidents involved organisational or sensitive material being uploaded to public AI tools.
& ``a huge volume of documents being uploaded into ChatGPT'' \blackcapsule{P12}\\

& Perceived severity and impact
& Although participants acknowledged incidents, many framed them as low severity due to the absence of catastrophic or customer-data-centric harm.
& ``real benign stuff'' \blackcapsule{P08}\\
\cline{2-4}

& Detection and organisational response
& Detection relied primarily on DLP and monitoring tooling, though participants acknowledged only partial visibility into informal AI use.
& ``to some extent'' \blackcapsule{P27}\\

\cline{2-4}

& Intent and awareness
& Across organisations, participants emphasised that breaches and near misses were driven primarily by a lack of awareness and well-intentioned behaviour rather than malicious intent.
& ``people were not trying to be malicious \ldots{} they were just asking questions'' \blackcapsule{P17}\\
\midrule

\cbtheme{19}{\textsc{\textbf{Controls \& Interventions}}}
& Prohibitive controls
& A small number of organisations adopted strict blocking of public AI tools, early in the generative AI adoption wave.
& ``blocked pretty early on in the surge of ChatGPT's popularity'' \blackcapsule{P04}\\
\cline{2-4}

 & Detective Controls (Monitoring and Visibility)
& Many organisations prioritised monitoring and follow-up over outright blocking, using visibility as the basis for education and response.
& ``our preference is not to block, it's to monitor, to understand, to educate, and respond'' \blackcapsule{P24}\\
\cline{2-4}

& Behaviour-Shaping Controls
& Soft interventions included nudges, feature restrictions, exceptions, and local conversations designed to shape behaviour without elimination.
& ``[department] leaders to have conversations with their teams'' \blackcapsule{P02}\\
\cline{2-4}

& Exception-Based Controls
& In some cases, organisations adopted exception-based approaches grounded in trust-based formal exception and risk-acceptance processes.
& ``then there needs to be an exception \ldots{} people \ldots{} accept the risk.'' \blackcapsule{P17}\\


& Substitution-based controls
& Providing sanctioned enterprise AI tools was seen as a practical way to curb shadow AI use without suppressing productivity.
& ``if you don't give them the tools, they'll find a way to get the tools'' \blackcapsule{P07}\\
\cline{2-4}

& Cultural and capability controls
& Training, awareness, and anti-heavy-handed governance were used to balance safe use with innovation and trust.
& ``stifle innovation'' \blackcapsule{P09} \\
\cline{2-4}

& Limits of Blocking as a Control Strategy
& Participants consistently described outright elimination as unrealistic given personal devices, embedded AI features, and simple workarounds.
& ``it's really hard to control, particularly with people using their own devices'' \blackcapsule{P05}\\
\midrule

\cbtheme{18}{\textsc{\textbf{Security, Compliance \& Regulatory Risks}}}
& Sensitive data exposure beyond organisational assurance boundaries
& Participants recognised sensitive-data exposure as a serious risk, even where some organisations expressed confidence in their internal controls.
& ``It's not that the data leaking is not important, it's just not nearly as important for us'' \blackcapsule{P17}\\
\cline{2-4}

& Insider-based exfiltration via generative AI interfaces
& Public AI tools were described as a low-friction pathway through which authorised users could move sensitive information outside approved security boundaries.
& ``You can log onto ChatGPT, copy and paste some customer data across \ldots{} and you'll have customers sitting in your history'' \blackcapsule{P08}\\
\cline{2-4}

& Loss of assurance from unassessed AI use in approved systems
& AI-mediated processing inside approved tools can expand sensitive inference and reliance without formal review, accountability, or auditability.
& ``what someone else is getting paid or some other private information for the employees'' \blackcapsule{P01}\\
\cline{2-4}

& AI-mediated access expansion from unapproved use cases
& AI-enabled aggregation and inference can expand effective access to sensitive information beyond intended disclosure boundaries.
& ``such as what someone else is getting paid or other private information'' \blackcapsule{P01} \\
\cline{2-4}

& Decision and epistemic integrity failure from unvetted model outputs
& Informal reliance on unvetted AI outputs created risks of automation bias, hallucination propagation, and weak accountability for downstream decisions.
& ``people are still accepting the answer \ldots{} because that's what Copilot or ChatGPT said'' \blackcapsule{P14}\\
\cline{2-4}

& Opaque third-party and supply-chain dependencies
& Routine shadow AI use silently introduced trust, dependency, and visibility risks tied to vendor-controlled models, infrastructure, and data practices.
& ``how our data is being used to further the interests of vendors versus our own interests'' \blackcapsule{P21}\\

\end{longtable}

\normalsize


\section*{Appendix B: Positioning Against Existing Shadow-AI Literature}
\label{app:related-work-comparison}

Table~\ref{tab:comparison_shadow_ai_gaps} positions this study against representative academic literature and industry reports on shadow AI. The comparison is structured around the main gaps identified in Section~\ref{sec:results}, including the extent to which prior work focuses explicitly on shadow AI, provides empirical organisational evidence, addresses CI or regulated contexts, examines organisational drivers and constraints, incorporates leadership perspectives and responses, and analyses governance and control practice in situ.

The table is not intended to provide an exhaustive systematic literature review. Instead, it highlights how the present study differs from representative strands of existing work. Prior literature has established shadow AI as an emerging governance and security concern, but much of it remains conceptual, prescriptive, survey-based, or focused on general organisational settings. In contrast, this study provides empirical organisational evidence from regulated CI environments and examines how shadow AI is shaped by operational demand, governance lag, organisational culture, and assurance constraints.

\begin{table*}
\centering
\caption{Comparison of our work with representative shadow-AI literature and industry reports, structured around key gaps in existing work. \lmark\ indicates that a dimension is addressed only partially or indirectly rather than as a sustained empirical or analytical focus.} 
\label{tab:comparison_shadow_ai_gaps}
\begin{tabular}{l c c c c c c c}
\toprule
\textbf{Ref.} & \textbf{Year} &
\begin{tabular}[c]{@{}c@{}}\textsc{Primary}\\\textsc{Focus on}\\\textsc{Shadow AI}\end{tabular} &
\begin{tabular}[c]{@{}c@{}}\textsc{Empirical}\\\textsc{Organisational}\\\textsc{Evidence}\end{tabular} 
&
\begin{tabular}[c]{@{}c@{}}\textsc{CI or}\\\textsc{Regulated}\\\textsc{Context}\end{tabular} 
&
\begin{tabular}[c]{@{}c@{}}\textsc{Organisational}\\\textsc{Drivers and}\\\textsc{Constraints}\end{tabular} 
&
\begin{tabular}[c]{@{}c@{}}\textsc{Leadership}\\\textsc{Perspectives}\\\textsc{and Responses}\end{tabular} 
&
\begin{tabular}[c]{@{}c@{}}\textsc{Governance}\\\textsc{and Controls}\\\textsc{in Practice}\end{tabular} \\
\midrule
\multicolumn{8}{l}{\textit{\textbf{Conceptual, governance, and prescriptive literature}}} \\
\midrule
\cite{van2024bring} & 2024 & \lmark & \cmark & \xmark & \lmark & \cmark & \lmark \\
\cite{puthal2025shadows} & 2025 & \cmark & \xmark & \lmark & \lmark & \xmark & \lmark \\
\cite{ross2025shadow} & 2025 & \cmark & \lmark & \lmark & \lmark & \lmark & \lmark \\
\cite{ogungbemi2025shadow} & 2025 & \cmark & \xmark & \cmark & \lmark & \xmark & \lmark \\

\midrule
\multicolumn{8}{l}{\textit{\textbf{Industry reports and surveys}}} \\
\midrule
\cite{recoshadowAI2025} & 2025 & \cmark & \lmark & \xmark & \xmark & \xmark & \xmark  \\
\cite{josys2025ShadowAI} & 2025 & \cmark & \lmark & \lmark & \lmark & \lmark & \lmark  \\
\cite{nrg2024outoftheshadows} & 2024 & \cmark & \cmark & \lmark & \cmark & \cmark & \cmark    \\
\midrule
\textbf{\textit{Ours}} & \textbf{2026} & \cmark & \cmark & \cmark & \cmark & \cmark & \cmark  \\
\bottomrule
\end{tabular}%
\end{table*}

Table~\ref{tab:comparison_shadow_ai_gaps} clarifies the contribution of the paper in relation to prior work. Whereas existing studies frequently identify data leakage, governance blind spots, and compliance risks, they provide limited insight into how shadow AI is perceived, governed, and mitigated by organisational decision-makers in highly regulated CI settings. This study addresses that gap by combining sector-specific interviews with a socio-technical analysis of assurance erosion.

\section*{Appendix C: Supplementary Conceptual Figures}
\label{app:supplementary-conceptual-figures}
This appendix section provides supplementary conceptual figures that support the framing and interpretation of the empirical findings. These figures are not intended to introduce additional empirical claims beyond those reported in Section~IV. Instead, they provide compact visual summaries of the risk categories, historical evolution, and assurance-erosion pathway discussed throughout the paper.


\subsection{Evolution from Shadow IT to Shadow AI}
\label{app:shadow-ai-evolution}

\begin{figure*}
    \centering
    \includegraphics[width=1\linewidth]{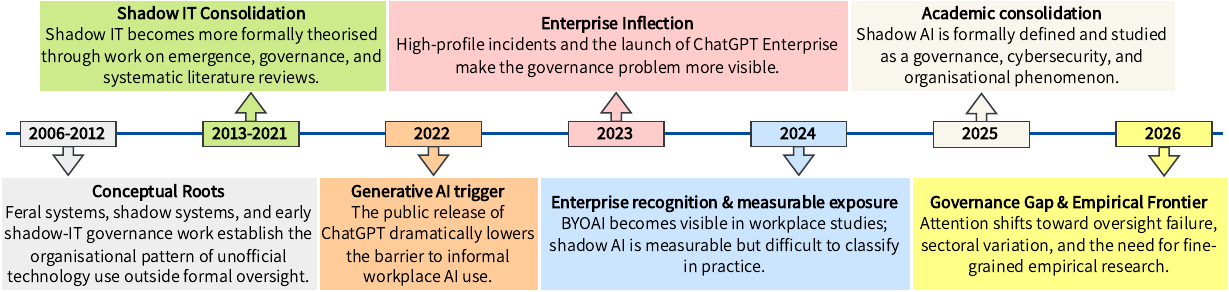}
    \caption{From shadow IT to shadow FAI: Evolution of shadow AI from conceptual roots to the current governance frontier.}
    \label{fig:timeline}
\end{figure*}

Figure~\ref{fig:timeline} summarises the conceptual evolution from early work on feral systems and shadow IT to contemporary concerns around shadow AI. The figure highlights that shadow AI did not emerge in isolation. Rather, it builds on a longer organisational pattern in which workers adopt unofficial technologies to address unmet operational needs. Generative AI, however, amplifies this pattern by increasing its scale, speed, and visibility.

The figure also supports the related-work discussion by illustrating how shadow AI both extends and diverges from earlier shadow-technology phenomena. Traditional shadow IT typically involved unofficial systems, applications, or infrastructure that could eventually be inventoried or governed as technical assets. Shadow AI, by contrast, frequently operates at the interaction layer through prompts, uploaded documents, AI-mediated synthesis, and embedded model capabilities. This makes the phenomenon harder to classify and reconstruct using conventional asset-centric governance approaches.

\subsection{Pathway from Shadow-AI Use to Assurance Erosion}
\label{app:assurance-erosion-pathway}
\begin{figure}
    \centering
    \includegraphics[width=0.8\linewidth]{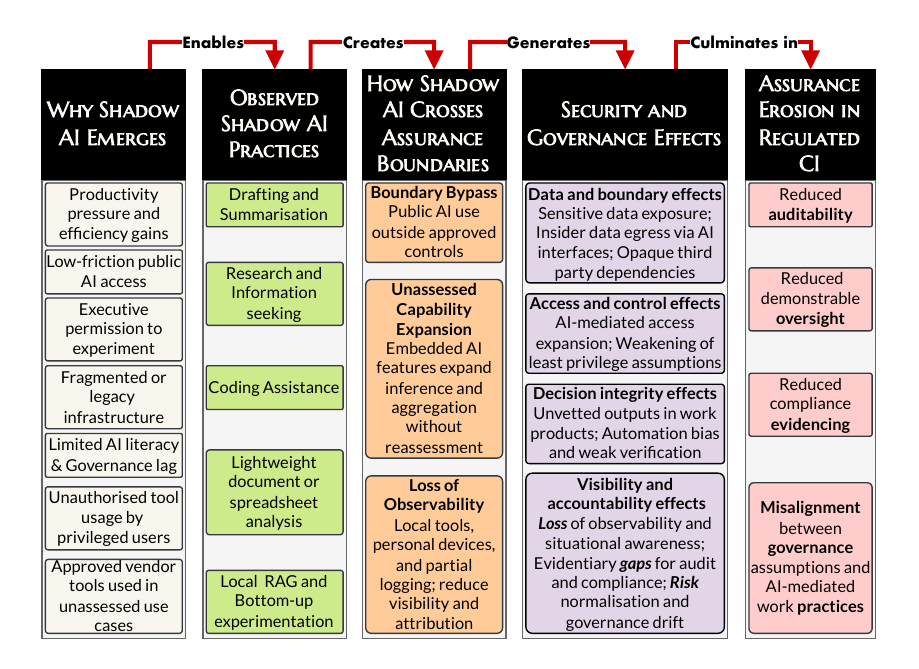}
    \caption{From shadow-AI use to assurance erosion in regulated CI. Routine, human-in-the-loop shadow-AI practices emerge from organisational pressures and cross assurance boundaries through boundary bypass, unassessed capability expansion, and loss of observability, generating cumulative security, governance, and compliance effects that culminate in assurance erosion.}
    \label{fig:result_summary}
\end{figure}

Figure~\ref{fig:result_summary} visualises how routine shadow-AI practices can produce cumulative assurance erosion in regulated CI environments. The pathway begins with organisational drivers, such as productivity pressure, low-friction public AI access, executive permission to experiment, fragmented infrastructure, limited AI literacy, and governance lag. These drivers enable observed practices such as drafting and summarisation, research and information seeking, coding assistance, lightweight document or spreadsheet analysis, and local RAG or bottom-up experimentation.

Figure~\ref{fig:result_summary} complements the threat model in Section~\ref{sec:dicussion} by showing how shadow AI risk emerges as a pathway rather than as a single failure event. The figure links observed drivers and practices to three assurance-boundary mechanisms: boundary bypass, unassessed capability expansion, and loss of observability. These mechanisms then generate data and boundary effects, access and control effects, decision-integrity effects, and visibility and accountability effects. The pathway culminates in reduced auditability, reduced demonstrable oversight, reduced compliance evidencing, and misalignment between governance assumptions and AI-mediated work practices.

\normalsize

\section*{Appendix D: Interview Questions}
\label{app:interview-questions}

The interview formed part of a broader semi-structured study on Frontier AI adoption, governance and risk management in CI organisations. Shadow AI was one of the main topics within this wider protocol. This appendix reports only the questions directly relevant to shadow AI, including questions concerning informal AI use, organisational awareness, risk assessment, perceived risk severity, and governance controls.

\subsection{Participant Demographics and Organisational Context}

\begin{itemize}

    \item What is your current role or title, and where does this role sit within your organisation's leadership structure?
    
    \item Which critical infrastructure sector does your organisation operate in?
    
    \item What is your primary professional background or discipline?
    
    \item What is the highest level of formal education you have completed?
\end{itemize}

\subsection{Organisational Understanding and Definition of Frontier AI}

\begin{itemize}
    \item How does your organisation understand or define Frontier AI?
\end{itemize}

\subsection{Frontier AI Adoption Context}

\begin{itemize}
    \item What are the primary barriers to the adoption of Frontier AI systems in critical infrastructure organisations?
\end{itemize}

\subsection{Shadow AI and Informal AI Use}

\begin{itemize}
    \item Has your organisation experienced any shadow or informal use of Frontier AI or other AI tools?
    
    \item Where do you think shadow AI is most common within your organisation? For example, is it concentrated in specific teams or functions, or is it more widespread?
    
    \item What kinds of AI tools and tasks are typically involved in shadow or informal AI use?
    
    \item What measures or controls do you currently have, or would like to have, to detect, manage, or discourage the shadow use of Frontier AI?
\end{itemize}

\subsection{Risk Awareness and Assessment}

\begin{itemize}
    \item How would you describe your organisation's awareness of Frontier AI risks, including risks arising from both formal deployments and shadow or informal use?
    
    \item Does your organisation conduct formal risk assessments for AI systems?
    
    \item If yes, how does your organisation specifically assess risks associated with Frontier AI models?
\end{itemize}

\subsection{Perceived Severity of Shadow AI Risks}

\begin{itemize}
    \item How would you assess and rank the following shadow AI risks by severity? Please discuss each risk in detail, including its implications, and provide a clear rationale for your prioritisation.
    
    \begin{itemize}
        \item Shadow AI can expose sensitive systems and data to breaches.
        
        \item Lack of confidence in whether AI-related security measures adequately safeguard organisational information.
        
        \item Employees using unapproved AI tools may bypass security protocols, increasing organisational risk.
        
        \item Difficulty enforcing AI policies may lead to regulatory and legal exposure.
        
        \item Third-party risk: external AI services may not meet organisational security standards, creating dependency and assurance risks.
        
        \item Compliance breaches under sector-specific legislation, such as the Security of Critical Infrastructure Act.
    \end{itemize}
\end{itemize}
%



\end{document}